\documentclass[twocolumn]{aastex61}
\usepackage{multirow}
\usepackage{subfigure}

\usepackage{tikz}

\submitjournal{ApJ}
\received{December 14th, 2017}
\accepted{May 14th, 2018}

\shorttitle{Scoring Yield Tables}
\shortauthors{Philcox, Rybizki \& Gutcke}

\begin{document}

\title{On the Optimal Choice of Nucleosynthetic Yields, IMF and number of SN\,Ia for Chemical Evolution modelling}

\correspondingauthor{Oliver Philcox}
\email{ohep2@cam.ac.uk}

\author[0000-0002-3033-9932]{Oliver Philcox}
\affiliation{Institute of Astronomy, Madingley Road, Cambridge, CB3 0HA, UK}
\affil{Max Planck Institute for Astronomy,
	K\"onigstuhl 17, D-69117 Heidelberg, Germany}

\author[0000-0002-0993-6089]{Jan Rybizki}
\affil{Max Planck Institute for Astronomy,
K\"onigstuhl 17, D-69117 Heidelberg, Germany}

\author[0000-0001-6179-7701]{Thales A. Gutcke}
\affil{Max Planck Institute for Astronomy,
	K\"onigstuhl 17, D-69117 Heidelberg, Germany}



\defcitealias{2010A&A...522A..32R}{RK10}
\defcitealias{2015MNRAS.451.3693M}{MC15}
\defcitealias{2003PASP..115..763C}{CH03}
\defcitealias{Rybizki}{RJR17}
\defcitealias{2017arXiv171004222G}{GS17}

\begin{abstract}
To fully harvest the rich library of stellar elemental abundance data available, we require reliable models that facilitate our interpretation of them. Galactic chemical evolution (GCE) models are one such set, and a key part of which are the selection of chemical yields from different nucleosynthetic enrichment channels, predominantly asymptotic giant branch (AGB) stars, Type Ia supernovae (SNe\,Ia), and core-collapse supernovae (CC-SNe). Here, we present a scoring system for yield tables based on their ability to reproduce proto-solar abundances within a simple parametrisation of the GCE modelling software \textit{Chempy}, which marginalises over galactic parameters describing simple stellar populations (SSPs) and interstellar medium physics. Two statistical scoring methods are presented, based on Bayesian evidence and leave-one-out cross-validation and are applied to five CC-SN tables; (a) for all mutually available elements and (b) for a subset of the 9 most abundant elements. We find that the yields used by Prantzos et al. (P18, including stellar rotation) and Chieffi \& Limongi (C04) best reproduce proto-solar abundances for the two cases, respectively. The inferred best-fit SSP parameters for (b) are $\alpha_\mathrm{IMF}=-2.45^{+0.15}_{-0.11}$ for the initial mass function high-mass slope and $\mathrm{N}_\mathrm{Ia}=1.29^{+0.45}_{-0.31}\times10^{-3}$\,M$_\odot^{-1}$ for the SN\,Ia normalisation, which are broadly consistent across tested yield tables. Additionally, we demonstrate how \textit{Chempy} can be used to dramatically improve elemental abundance predictions of hydrodynamical simulations by plugging tailored best-fit SSP parameters into a Milky Way analogue from Gutcke \& Springel. Our code, including a comprehensive tutorial, is freely available and can additionally provide SSP enrichment tables for any combination of parameters and yield tables.

\end{abstract}


\submitjournal{ApJ}



\section{Introduction} \label{sec:intro}
In order to understand the chemical enrichment of galaxies and interpret the growing wealth of observations contributing to the elemental abundance space, we require accurate models which are able to trace the chemical content of the interstellar medium (ISM) as a function of cosmic time, taking into account chemical, physical and dynamical processes \citep{2003ceg..book.....M}. At the heart of almost every model including chemical evolution, are stellar enrichment processes that are integrated over the stellar initial mass function (IMF) and used to predict the elemental feedback of a simple stellar population (SSP). 

For the most important (in terms of bulk metal contribution) stellar enrichment channels, namely CC-SN, SN\,Ia and AGB stars, there exists a vast menagerie of published yield tables with substantial differences, as shown for CC-SN and AGB yields in \cite{2010A&A...522A..32R,2015MNRAS.451.3693M} \citepalias[hereafter][]{2010A&A...522A..32R,2015MNRAS.451.3693M}. Even for SN\,Ia yields the discrepancies between authors have recently seen an increase due to the availability of 3D simulations \citep{2013MNRAS.429.1156S}, an effect which is likely to be emulated with the other nucleosynthetic channels \citep{2016PASA...33...48M}. Similarly, the most important SSP parameters, i.e. the high-mass slope of the IMF and the incidence of SN\,Ia \citep{2016ApJ...824...82C,Rybizki} (hereafter \citetalias{Rybizki}), have a substantial spread of literature values \citep[tab.7]{2016ApJ...824...82C}.

Despite these crucial uncertainties, recent cosmological hydrodynamical simulations have produced increasingly realistic abundance distributions \citep[e.g.][]{2012MNRAS.424L..11F,Naiman}. In order to improve upon these results and include more elements in the analysis, parameter studies to determine the best combination of yield set and SSP parameters need to be carried out. Hydrodynamical simulations are computationally far too expensive to use for such searches \citep{2014MNRAS.442.3745M}, therefore it is common to rely upon geometrically simplified galactic chemical evolution (GCE) models assuming a well-mixed ISM.

Previous studies have investigated the effect of varying the SSP parameters \citep[e.g.][]{2005A&A...430..491R,2015MNRAS.449.1327V,2016ApJ...824...82C}, the employed yields (e.g. \citealt{2002IAUS..187..159G}; \citetalias{2010A&A...522A..32R}) or both (e.g. \citealt{1997MNRAS.290..471G,2011MNRAS.414.3231K}; \citetalias{2015MNRAS.451.3693M}; \citealt{2017ApJ...835..224A}) on the predicted abundance distributions, but usually only one parameter at a time is left free or if not, only a single set of yields is considered. From such analysis no firm conclusion can be drawn on the optimal parameter and/or yield set since restricting to a single set of yields (and likewise to fixed parameter values) has been shown to produce significantly biased results (\citealt{1998ApJ...501..675G}; \citetalias{Rybizki}).

In the literature, the two main investigations into yield set comparison are \citetalias{2010A&A...522A..32R} and \citetalias{2015MNRAS.451.3693M}. Both test different AGB star and CC-SN yield tables against an extensive set of observables using a fiducial GCE model with radial zones. Whereas \citetalias{2010A&A...522A..32R} analyse many different elemental abundances and discuss in depth observational uncertainties and the effects of nucleosynthetic modelling assumptions, they lack a measure for goodness of fit. \citetalias{2015MNRAS.451.3693M} improves upon this by employing a $\chi^2$ statistic on their binned observational constraint. The drawback of this is that outliers are penalised disproportionately, preventing them from including elements with less well established yields. Even though \citetalias{2010A&A...522A..32R} \citepalias{2015MNRAS.451.3693M} test two (six) IMFs they hold all other GCE parameters fixed during their yield set comparison. This severely underestimates both degeneracies with the yield tables, and uncertainties in the best yield table determination arising from GCE parameters \citep{2016ApJ...824...82C}.

In this paper we present a flexible technique capable of generating scores for nucleosynthetic yields only relying on a single well-established observational constraint; the solar elemental abundances \citep{2009ARA&A..47..481A}. We present two different statistical metrics and apply them to five CC-SN tables using (a) all available elements and (b) the subset of the nine most abundant ones. The flexible GCE model \textit{Chempy} \citepalias{Rybizki} is used as a framework and sped up by means of neural networks such that we are able to properly marginalise over nuisance chemical evolution parameters for the first time in such an analysis. To account for uncertainties in the model and yield tables, we include a variable error model whose shape parameter is also marginalised out. Thus, this technique can determine the best combination of yields and, as a side effect, provide best-fit SSP parameters for any given set of yields and elements. To demonstrate this functionality we predict best-fit values for the IMF high mass slope and SN\,Ia normalization for a Milky Way-like magneto-hydrodynamical simulation \citep{2017arXiv171004222G} using the {\sc AREPO} code \citep{2010MNRAS.401..791S}, under the constraint of a fixed yield set. The predicted parameters clearly improve the resulting abundance distribution of the simulation, as can be seen in figure \ref{fig:TNGvsAPOGEE}.

 Our analysis can be extended to other observations predicted by chemical evolution models (e.g. additional stellar elemental abundances, age-metallicity relations, halo gas metallicity or metallicity distribution functions). It can also be applied to score the yield tables of other chemical enrichment channels and is intended to guide modellers of nucleosynthetic yields who wish to test their own tables coming from different parameter studies (e.g. rotation vs.~static, rapid vs.~delayed detonation, mass-cut; \citealt{2002A&A...390..561M,2016ApJS..225...24P,2012ApJ...749...91F}) against observational constraints of GCE models. An additional feature of the code is the ability to produce SSP yield tables (net yield as a function of time and metallicity) for any given set of yields and SSP parameters similar to the module presented in \cite{2017arXiv171109172R}. 
 
 A Python implementation is freely available online \citep{Philcox}.\footnote{\url{https://github.com/oliverphilcox/ChempyScoring}} This includes a tutorial on implementing new yield tables, changing the set of elements and free chemical evolution parameters, training the neural network, running the Markov Chain Monte Carlo (MCMC) analysis, and computing both scores. The authors are happy to assist with this process upon request. 

We begin with a discussion of the \textit{Chempy} model and its study-specific modifications in section \ref{sec:chempy}. Section \ref{sec:obsData} describes the observational constraints along with the choice of implemented yield tables and elements, then the two statistical scores are presented in section \ref{sec:scores}. Results for the different CC-SN yield tables and element subsets are explored in section \ref{sec:results}. In this section we also demonstrate the practicality of our model by using our parameter predictions in a galaxy formation simulation. We conclude with a summary in section \ref{sec:conc}.

\section{Specifics of the GCE model}\label{sec:chempy}

\subsection{Implementation of \textit{Chempy}}\label{subsec:ChempyMods}
A complete discussion of the \textit{Chempy} model can be found in \citetalias{Rybizki}. For convenience we recapitulate the main characteristics in the following section. \textit{Chempy} is a simple one-zone GCE model which computes the evolution of a localised region of a well-mixed ISM throughout cosmic time. It is an open-box model which self-consistently incorporates both primordial infall and a self-enriching gas halo via a fixed outflow fraction and infall constrained by the gas required to sustain the star formation. 
Using input parameters describing SSP and ISM physics (parametrising the IMF, SN\,Ia rate, star formation rate (SFR, modelled as a gamma function), star formation efficiency (SFE) and the gas outflow fraction) together with hyperparameters (e.g.~nucleosynthetic yields), \textit{Chempy} can produce predictions, primarily of the elemental abundances of the ISM over time. By comparing mock observations derived from these to real data, we may construct a likelihood function, which can be sampled within a Bayesian framework via the Markov Chain Monte Carlo (MCMC) routine \texttt{emcee} \citep{2013PASP..125..306F}, thereby constraining the input parameters. 
Here, the key functionality of \textit{Chempy} is that of a ``black-box", producing a set of model elemental abundances at the time of solar birth for input parameter vector $\theta$. 

In this study, the \textit{Chempy} free parameters $\theta$ (table \ref{tab:priors}) have been updated (cf. \citetalias[table 1]{Rybizki}), changing the IMF parametrisation to that of \citet[table 1]{2003PASP..115..763C} (hereafter \citetalias{2003PASP..115..763C}) with variable high-mass slope $\alpha_\mathrm{IMF}$ and considering the SFR peak parameter in logarithmic form. In addition, the formerly used corona mass factor, $\log_{10}(f_\mathrm{corona})$, is fixed to its prior value of 0.30, since it was found to be degenerate with the outflow fraction $\mathrm{x_{out}}$. Also, the SN\,Ia delay time distribution parameter $\log_{10}(\tau_\mathrm{Ia})$ is set to $-0.80$, since this cannot be meaningfully constrained by proto-solar abundances alone. A Gaussian prior is assumed for all 5 \textit{Chempy} parameters with mean and standard deviation as given in table \ref{tab:priors}.

\begin{tiny}
\begin{table*}
\begin{minipage}{\textwidth}
\begin{center}
\caption{Free \textit{Chempy} and model error parameters used in this study, with their prior values and Gaussian widths.}
\begin{tabular}{ rl|ccc }
$Parameter$ & Description & $\overline{\theta}_\mathrm{prior}\pm\sigma_\mathrm{prior}$ & Limits & Approximated prior based upon: \\
 \hline
  \multicolumn{5}{c}{\textit{Global stellar (SSP) parameters}}\\
\hline
  $\alpha_\mathrm{IMF}$ & High-mass slope of the \citetalias[tab.1]{2003PASP..115..763C} IMF & $-2.3\pm0.3$ & $[-4,-1]$ & \citetalias[tab.1]{2003PASP..115..763C} \\
  $\log_{10}\left(\mathrm{N}_\mathrm{Ia}\right)$ & Number of SN\,Ia exploding per M$_\odot$ over 15\,Gyr & $-2.75\pm0.3$ & $[-5,-1]$ & \citet[tab.1]{2012PASA...29..447M}\\
\hline
  \multicolumn{5}{c}{\textit{Local ISM parameters}}\\
\hline
  $\log_{10}\left(\mathrm{SFE}\right)$ & Star formation efficiency governing the infall and ISM & $-0.3\pm0.3$ & $[-3,2]$ & \cite{2008AJ....136.2846B}\\
  $\log_{10}\left(\mathrm{SFR}_\mathrm{peak}\right)$ & SFR peak in Gyr (scale of $\gamma$-distribution with $k=2$)& $0.55\pm0.1$ & $[-1,1]$ & \citet[fig\,4b]{2013ApJ...771L..35V}\\
 x$_\mathrm{out}$ & Fraction of stellar feedback outflowing to the corona & $\phantom{-}0.5\pm0.1$ & $[0,1]$ & \citetalias[tab.1]{Rybizki}\label{tab:priors}\\
\hline
 \multicolumn{5}{c}{\textit{Error model parameter (section \ref{subsec:errors})}}\\
\hline
$\log_{10}\beta$ & Beta distribution shape parameter (fig\,\ref{fig:beta}) & $1.0\pm0.5$ & [0,$\infty$) & Initial studies at fixed $\beta$
\end{tabular}
\end{center}
\end{minipage}
\end{table*}
\end{tiny}

The simulation time-step was increased by a factor of five to 0.5 Gyr, since this was found to give sufficiently precise results in much reduced computation time. Furthermore, before running \texttt{emcee}, an initial parameter-space optimisation is now performed via the \texttt{scipy} package \citep{Scipy} allowing for faster convergence. To prevent unrealistically low SFRs we impose the constraint that the SFR at the time of the Sun's birth should be at least 5\% of the median value, i.e. 5\% of the SFR if the SFR was constant\footnote{This is merely to ensure that the SFR parameter does not peak at very early times which would result in practically no SFR at the solar birth. The code would otherwise still provide ISM abundances in this case which will generally be higher and therefore biased.} 

\subsection{Error Parametrisation}\label{subsec:errors}
In previous versions of \textit{Chempy} no account has been made for model errors that derive from a number of sources including modelling assumptions, yield table inaccuracies and missing stellar enrichment channels. The effect of this is that a few badly reproduced elements (e.g. K in \citetalias[figure 14]{Rybizki}) strongly dominate the likelihood and therefore bias the posterior parameter distribution. 


Here we introduce a model error distribution, $f(\sigma_\mathrm{m})$, which is marginalised over, treating each element equally for simplicity. An associated shape parameter is used to favour large or small model errors and is integrated over. This clearly does not fully account for the different errors in each element, but is able to make first-order corrections and mitigate biases arising from poorly predicted elements.

Each element in the observational dataset $\mathcal{O}_\mathrm{s}$ (of size $n_\mathrm{el}$) has abundance $\mathcal{O}_{\mathrm{s},i}$ and error $\sigma_{\mathrm{obs},i}$ which can be compared to the predicted value, $d_{\mathrm{s},i}$, and model error, $\sigma_\mathrm{m}$, using
\begin{equation}
\mathcal{L}(\sigma_\mathrm{m}) = \prod_{i=1}^{n_\mathrm{el}}(2\pi\sigma_i^2)^{-0.5}\exp{\left(-\frac{(\mathcal{O}_{\mathrm{s},i}-\mathrm{d}_{\mathrm{s},i})^2}{2\sigma_i^2}\right)},
\end{equation}
(cf. \citetalias[equation 9]{Rybizki}) for combined error $\sigma_i^2 = \sigma_{\mathrm{obs},i}^2 + \sigma_\mathrm{m}^2$.

A beta function was chosen for the error parametrisation with probability density function (PDF)
\begin{equation}
f(\sigma_\mathrm{m};\alpha,\beta) = \frac{\Gamma(\alpha+\beta)}{\Gamma(\alpha)\Gamma(\beta)}\sigma_\mathrm{m}^{\alpha-1}(1-\sigma_\mathrm{m})^{\beta-1},
\end{equation}
for $0<\sigma_\mathrm{m}<1$, with shape parameters $\alpha$ (set to unity) and $\beta$, where $\Gamma(z)$ is the gamma function. 

\begin{figure}[htb]
\includegraphics[width=\linewidth]{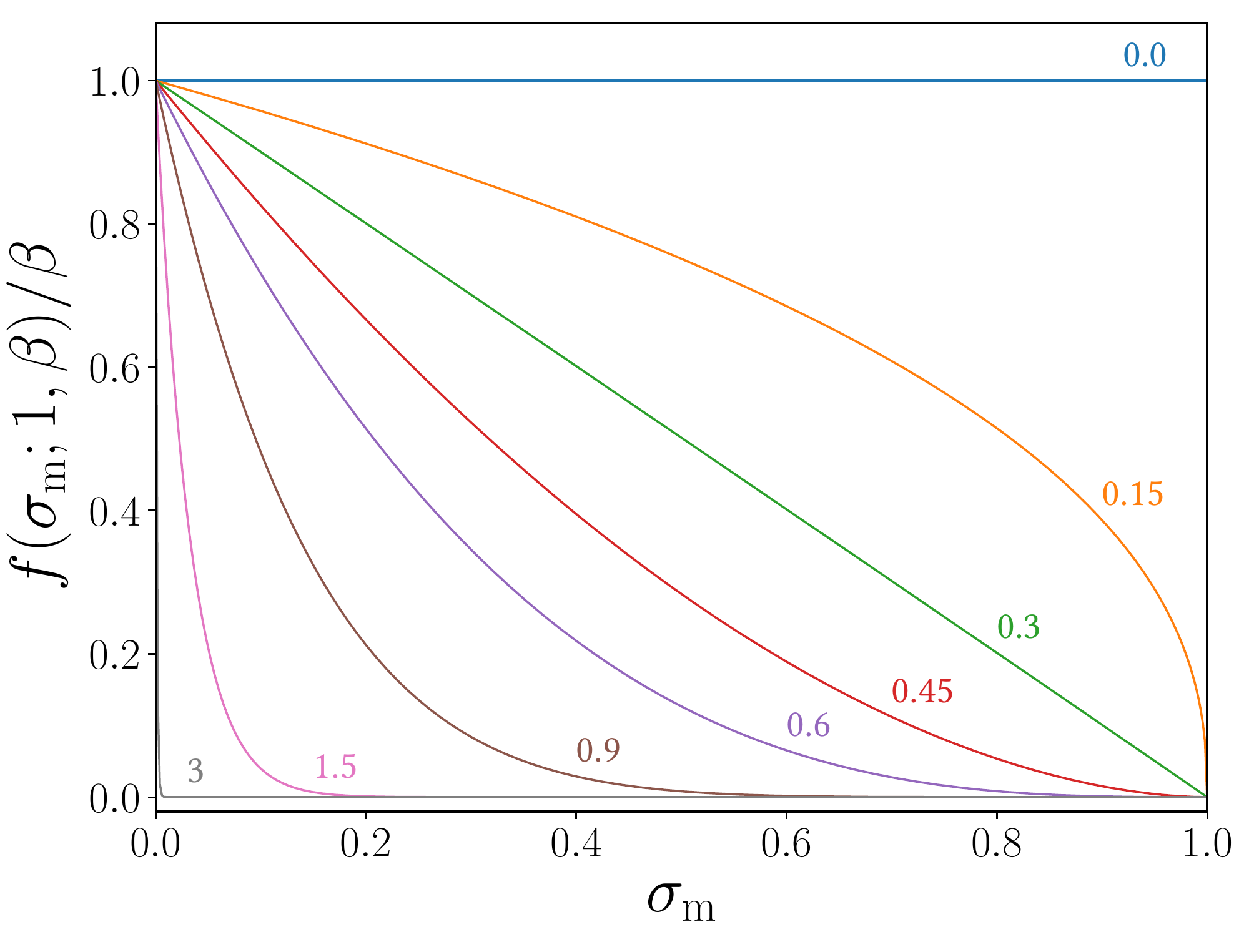}
\caption{Beta distribution PDF, $f(\sigma_\mathrm{m}; 1,\beta)$, for marked values of $\log_{10}\beta$. Curves are rescaled to have a peak of unity, and the limiting case of $\beta \rightarrow 0$ ($\infty$) represents a flat (zero) error within [0,1] dex. This distribution is used as a model error for \textit{Chempy} and we leave $\beta$ as a free parameter in this analysis.}\label{fig:beta}
\end{figure}

By varying $\beta\in {[1,\infty})$ we can change the error tolerance, including the limiting cases of uniform and zero model error, as shown in figure \ref{fig:beta}. The likelihood at fixed $\beta$, $\mathbb{L}(\beta)$, is then computed using
\begin{equation}\label{eq:marg-likelihood}
\mathbb{L}(\beta) = \int_0^1 f(\sigma_\mathrm{m};1,\beta)\mathcal{L}(\sigma_\mathrm{m})\mathrm{d}\sigma_\mathrm{m},
\end{equation}
approximated via numerical integration over a grid of $\sigma_\mathrm{m}$ evenly spaced in [0,1]. We allow the shape parameter to vary freely inside the aforementioned limits, with larger values of $\beta$ indicating a preference of the model for smaller errors.

\subsection{Neural Networks}\label{subsec:Neural}
Creation of a yield table score requires many computations of the posterior. In order to speed up the calculation, a neural network is implemented to predict the abundance output for a given set of input parameters, $\theta$, practically acting as a fast interpolator. We make use of the PyTorch software,\footnote{\url{pytorch.org}} creating a simple network with 5 inputs ($\theta$, excluding $\beta$) and $n_\mathrm{el}$ outputs ($\mathrm{d}_s$).

The training set for the network is a grid of $10^5$ points, with 10 values of each parameter drawn from an inverse Gaussian distribution spanning $2.8\sigma_\mathrm{prior}$ in parameter space (such that all points are inside the prior limits). Network hyperparameters were optimised using an independent dataset to reduce bias, giving a 30 neuron network with 1 hidden $\tanh$ layer \citep{LeCun:1998:EB:645754.668382,Stanford}. An Adam optimiser \citep{2014arXiv1412.6980K} with learning rate 0.007 is used to train the network over 5000 epochs with the L1 loss function.

The network errors (discussed in appendix \ref{appenA}) are negligible compared to those of the observational dataset, and the overall effect of the implementation is to reduce computation time from 600 ms to 7 ms for each posterior evaluation, although we note that the network must be retrained for each new yield set.

\section{Data and yield table selection}\label{sec:obsData}

\subsection{Stellar Abundance Data}\label{subsec:ObsData}
To assess the likelihood of model parameters, we must compare \textit{Chempy}'s predictions with observed stellar abundances. Modern spectral libraries \citep[e.g. APOGEE: ][]{2017AJ....154...94M} provide many such examples. Clearly, the model complexity increases with the number of data-points, as ISM parameters must be allowed to vary between stars (cf. \citetalias[figure 16]{Rybizki}). This vastly inflates the number of free parameters and we must also consider errors in the age of each star, adding further computational expense. Here, we adopt the simplest possible prescription, using only proto-solar abundances to assess how this can constrain the choice of yield table.

$\mathrm{[X/Fe]}_\odot$ and $\mathrm{[Fe/H]}_\odot$ solar abundances are used from \cite{2009ARA&A..47..481A}, as independently verified by analysis of meteorites and comets \citep{1989Icar...80..225L,2009LanB...4B...44L}. We apply the heavy element corrections of \cite{2002JGRA..107.1442T} to convert these to proto-solar abundances, adding 0.01 (0.04) dex to the [He/Fe] ([Fe/H]) abundance and increasing the uncertainties by 0.01 dex for each data-point. These are compared to \textit{Chempy}'s predictions at time $\mathrm{t_p} - \tau_\odot$ for current time $\mathrm{t_p}$ and solar age $\tau_\odot = 4.5$ Gyr \citep{1999A&A...343..990D}, assuming negligible age error. No further observational constraint is used in our analysis.

\subsection{Yield Tables}\label{subsec:yields}
Nucleosynthetic yields are implemented for three main processes; AGB feedback, CC-SN and SN\,Ia, using a mass range of $0.5 - 8$ ($8 - 40)$ $\mathrm{M}_\odot$ for AGB (CC-SN) events \citep{2009ARA&A..47...63S}. For higher stellar masses up to the IMF limit of $100$ $\mathrm{M}_\odot$, a new process was implemented to simulate the effects of ``failed supernovae" \citep{2017MNRAS.469.1445A}, returning 75\% of the initial stellar composition to the ISM via winds without enrichment, leaving the remainder to supposedly form a black hole. This increases the mass in remnants by $\sim6\%$ and decreases the alpha enhancement by $\sim35\%$. It also reduces extrapolation error, since the CC-SN yield tables we use do not calculate explosions for stars above $40 \mathrm{M}_\odot$ (see table \ref{tab:yieldsets}).\footnote{For a comparison between models with a CC-SN mass limit of 40~$M_\odot$ (including failed supernovae) and 100~$M_\odot$ (with no failed supernovae) see tables \ref{tab:ism_parameters}(c) and \ref{tab:ism_parameters}(d).}

To provide a diagnostic test of the statistical analysis, we compare five CC-SN yield tables, listed in table \ref{tab:yieldsets}, along with the default \cite{2010MNRAS.403.1413K} (AGB) and \cite{2013MNRAS.429.1156S} (SN\,Ia) yields from \citetalias[table 2]{Rybizki}. Here, logarithmic interpolation is used to compute the yields for any given metallicity.\footnote{Rerunning a test yield set for linear interpolation instead shows that this does not affect the analysis; comparative scores are consistent within their stated errors and all posterior predictions are consistent within $0.2\sigma$.} Some element of bias will be introduced by imperfections in the yield sets that are held constant in the study, although these are fairly well understood compared to CC-SN yields and their nucleosynthetic contribution is much smaller.

Only net yields are used here, such that the tables provide newly synthesized material and the remainder comes from the initial SSP composition. This requires that the C04 and W17 yields, originally presented in gross format, are modified to convert them into net yields utilising the relevant initial abundances; \cite{1989GeCoA..53..197A} solar abundances and \cite{2013ApJ...774...75W} abundances for the two models respectively. In addition, we note that the N13 yields are based on those of \cite{2006ApJ...653.1145K} (excluding their hypernova yields), and the currently unpublished W17 yields are those used in \cite{2016MNRAS.463.3755C}. The R17 yields come in two detonation flavours; `rapid' and `delayed' \citep{2012ApJ...749...91F} of which we use the latter.\footnote{'Rapid' detonations were also considered, but their score was found to be extremely low.} The P18 yields are those of \citet{2018arXiv180509640L}, averaged over three rotation speeds using the metallicity dependent rotation distribution prescription described in \citet[fig.\,4]{2018MNRAS.tmp..302P}. None of the other tested CC-SN yields include rotation in their calculation and all but C04 include stellar winds.

\begin{table}
\caption{CC-SN yield tables compared in this study and their mass and metallicity ranges.}
\centering
\begin{tabular}{c|c|c|c}
Abbr. & Yield Table& Masses & Metallicities\\
\hline
C04 & \cite{2004ApJ...608..405C}&[13,35]&[0,0.02]\\
N13 & \cite{Nomoto}&[13,40]&[0.001,0.05]\\
W17 & West \& Heger (in prep.)&[13,30]&[0,0.3]\\
R17 & \cite{Ritter}&[12,25]&[0.0001,0.02]\label{tab:yieldsets}\\
P18 & \cite{2018arXiv180509640L}\footnote{We use the rotation parametrisation of \cite{2018MNRAS.tmp..302P} for these yields.}&[13,120]&[0.0000134,0.0134]
\end{tabular}
\end{table}

\subsection{Choice of Elements}\label{subsec:elementchoice}
The statistical methods below can be used to calculate scores for any combination of elements. The implemented CC-SN tables provide yields for all elements up to Ge, but we exclude Li, Be and B as these abundances are significantly altered by unmodelled nucleosynthetic pathways. This gives a total of 29 tracked elements resulting in 28 predictions ($n_\mathrm{el} = 28$) as H is only implicitly included via normalisation in the abundances. No statistical weighting is applied to the other elements, though the Fe prediction affects each data point via the chosen [X/Fe] normalisation. 


To demonstrate how the resulting scores depend on the chosen elements, we also apply our analysis to the main set of elements used in hydrodynamical simulations \citep[e.g.][]{2010MNRAS.401..791S}, which are the 9 most abundant by mass fraction (H, He, C, N, O, Ne, Mg, Si, Fe), giving $n_\mathrm{el} = 8$. Adaptation to other sets of elements is straightforward, and is described in the online tutorial. 

\section{Objective scoring models}\label{sec:scores}
Here we propose two complimentary methods to calculate an objective yield table score, discussing their implementation and limitations.

\subsection{Bayes Factor approach}\label{subsec:bayes}
To compare yield sets $\chi$ given observations $\mathcal{O}_\mathrm{s}$, we may simply calculate the ratio of their Bayesian evidences (e.g. \citealt{10.2307/2291091,2012A&A...546A..89B});
\begin{equation}\label{eq:Bayes}
\mathsf{K}_\mathrm{B} = \frac{\mathrm{P}(\mathcal{O}_\mathrm{s}|\chi_1)}{\mathrm{P}(\mathcal{O}_\mathrm{s}|\chi_2)}.
\end{equation}
This is the classical Bayes factor, and if $\mathsf{K}_\mathrm{B} \gg 1$ we may conclude that $\chi_1$ is the better predictor of $\mathcal{O}_\mathrm{s}$. The evidence is found via integration;
\begin{eqnarray}\label{eq: evidence}
\mathrm{P}(\mathcal{O}_\mathrm{s}|\chi_i) &=& \int \mathrm{P}(\mathcal{O}_\mathrm{s}|\theta,\beta,\chi_i)\mathrm{d}^5\theta \mathrm{d}\beta \nonumber\\
&=& \int\mathbb{L}(\mathcal{O}_\mathrm{s}|\theta,\beta,\chi_i)\mathcal{P}(\theta,\beta)\mathrm{d}^5\theta \mathrm{d}\beta,
\end{eqnarray}
for posterior $\mathrm{P}(\mathcal{O}_\mathrm{s}|\chi_i)$, likelihood $\mathbb{L}(\mathcal{O}_\mathrm{s}|\theta,\beta,\chi_i)$, and combined prior $\mathcal{P}(\theta,\beta)$ (table \ref{tab:priors}), with the domain taken as the entire permitted parameter range. We marginalise over $\beta$ to take into account all error model shapes, assuming a broad Gaussian prior with mean and width estimated from initial studies at various fixed values of $\beta$.

The posterior is tightly peaked in parameter space about the median parameter values (shown in figure \ref{fig:posteriorPlot} for the two limiting values of $\beta$) and can thus be fairly well fit using a multivariate Gaussian (with covariances drawn from the MCMC PDF output). The parameter space integration, i.e, calculation of the Bayesian evidence, can then be performed efficiently via a Monte Carlo importance sampling routine \citep{2002nrca.book.....P} implemented with the \texttt{scikit-monaco} Python package.\footnote{Bugnion 2013, \url{http://scikit-monaco.readthedocs.io/}} Accuracy of the method has been checked using a more rudimentary (and slower) Monte Carlo method via the \texttt{mcint} package.\footnote{Snowsill 2011, \url{http://pypi.python.org/pypi/mcint/}} A numerical precision of 0.5\% was obtained using 50,000 posterior samples, which takes around 5 minutes on a modern 8-core machine. 

As noted in appendix \ref{appenA}, the neural network has increasing error towards the edges of parameter space, itself truncated by the finite prior domains (table \ref{tab:priors}). Computation of equation \ref{eq: evidence} for increasingly broad parameter ranges shows that there are negligible contributions from regions outside $3\sigma_\mathrm{prior}$, validating our approximations. 

We can thus obtain the Bayesian evidence, henceforth denoted by $\mathcal{S}_\mathrm{B}$, for any combination of yield sets, although this method retains significant dependence on the prior, $\mathcal{P}$, which is set using reasonable broad distributions (see table \ref{tab:priors}).

\subsection{Cross-Validation Approach}\label{subsec:cross-val}

An alternative scoring method for yield tables is considered employing the technique of leave-one-out cross-validation (LOO-CV) \citep[e.g.][]{Vehtari,2012A&A...546A..89B}. Using \textit{Chempy} and \texttt{emcee} we can predict the posterior PDF for the six f ree parameters, using only $n_\mathrm{el}-1$ of the $n_\mathrm{el}$ available data-points. Each set of posterior parameters in the PDF is then fed into \textit{Chempy} to produce a model elemental abundance PDF for the excluded element, which can be well approximated by a Gaussian.

Using the predicted Gaussian mean, $\mu_i$, and standard deviation, $\sigma_{\mathrm{pred},i}$, we may construct the $i$-th element likelihood;
\begin{equation}\label{eq:LOO-CV likelihood}
\mathcal{L}_i^{\mathrm{CV}} = (2\pi\sigma_i^2)^{-0.5}\exp\left(-\frac{(\mu_i-\mathcal{O}_{\mathrm{s}_i})^2}{2\sigma_i^2}\right)
\end{equation}
which compares the optimal element prediction with the observational abundances using the combined error $\sigma_i^2 = \sigma^2_{\mathrm{pred},i} + \sigma^2_{\mathrm{obs},i}$. This is iterated over all elements giving an overall score 
\begin{equation}\label{eq: LOO-CV score}
\mathcal{S}_\mathrm{CV}(\chi) = \left[\prod_{i=1}^{n_\mathrm{el}}\mathcal{L}_i^{\mathrm{CV}}(\chi)\right]^{1/n_\mathrm{el}},
\end{equation}
taking the geometric mean for normalisation, allowing comparison between scores with different $n_\mathrm{el}$. Here the shape parameter $\beta$ affects the results indirectly through broadening of the MCMC PDF. As before, yield tables are compared using the relative score $\mathsf{K}_\mathrm{CV} = \mathcal{S}_\mathrm{CV}(\chi_1) / \mathcal{S}_\mathrm{CV}(\chi_2)$, which can be computed in around half an hour on a modern 8-core machine for $n_\mathrm{el} = 28$. 


\section{Results} \label{sec:results}

Here we discuss the results of applying MCMC and the above metrics to the CC-SN tables defined in section \ref{subsec:yields}, with both sets of elements (section \ref{subsec:elementchoice}). In addition, section \ref{subsec:TNGYields} shows how the code can be used to constrain the IMF slope and SN\,Ia normalisation parameter for galaxy formation simulations, using {\sc AREPO} as an example. 

\subsection{Using All Available Elements}\label{subsec:AllResults}

\subsubsection{Posterior Parameter Distributions}\label{subsubsec:MCMCall}
Table \ref{tab:posteriors}\,(a) shows the results of running an MCMC analysis on the models with each choice of CC-SN yields for all available elements. $\log_{10}\beta$ is well constrained, with a peak value between $0.5$ and $0.8$, significantly different from both the prior and zero (the latter would indicate a flat error model). Larger $\beta$ favours smaller model errors, so from this alone one could surmise that the P18 and C04 yields best represent the data, although this does not take into account the full complexities of parameter space incorporated in the Bayesian score.

Meaningful constraints can only be placed on the SSP parameters. The ISM parameters, treated as nuisance parameters in this study, closely reproduce the priors as can be seen in table \ref{tab:ism_parameters} in appendix \ref{appenB}. This is unlike the constraints found in \citetalias{Rybizki} and results from the newly introduced model error which `flattens' the posterior PDF (cf.~figure \ref{fig:posteriorPlot}, where the lower plot shows the behaviour in the absence of an error model). The median of $\alpha_\mathrm{IMF}$ is mostly consistent with the fiducial value of $-2.3$, except more tightly constrained. Differences to the results of \citetalias[table 4]{Rybizki} arise from the addition of the ``failed supernova" routine, which gives less CC-SN enrichment, and also from the different IMF parametrisation used in this study. The W17 $\alpha_\mathrm{IMF}$ median posterior is anomalously high, most likely due to the large fraction of black holes produced by their supernova model.

The SN\,Ia normalisation, $\mathrm{N_{Ia}}$, is consistent with 10$^{-3}$~$\mathrm{M}_\odot^{-1}$ and a little lower than our prior, with C04 favouring a lower SN\,Ia rate than N13, matching the behaviour seen in \citetalias{Rybizki}. This underlines the possibility for parameters to be fine-tuned (within the ranges compatible with other observational constraints) for a chosen yield set in order to best match proto-solar abundances, as will be demonstrated in section \ref{subsec:TNGYields}.

	\begin{table*}
		\begin{minipage}{\textwidth}
			\begin{center}
				\caption{Inferred \textit{Chempy} and model error parameters (16, 50 and 84 percentiles of $\theta_\mathrm{posterior}$) obtained from an MCMC analysis for each yield set. Overall scores are obtained by running the Bayesian and LOO-CV analysis on a 6-dimensional parameter space consisting of 5 \textit{Chempy} free parameters and the error shape parameter $\beta$. Error estimates for $\mathcal{S}_\mathrm{CV}$ are as described in the text. The error for $\mathcal{S}_\mathrm{B}$ from the integration via importance sampling is negligible and other sources of error are discussed in the text. The ISM parameters (here treated as nuisance parameters) can be found in appendix \ref{appenB}, table \ref{tab:ism_parameters}, as well as the maximum posterior values for each run (which correlate mildly with the scores). The yield set termed TNG uses the combination of yields described in \citet[table 2]{Pillepich} with an upper mass limit for CC-SN explosions of 100\,M$_\odot$ (the default of IllustrisTNG, contrary to the 40\,M$_\odot$ used in this work).}
				\begin{tabular}{c|c|cc||cc|cc|cc}
					Yield set&\multicolumn{3}{c||}{Scores}&\multicolumn{2}{c|}{Error parameter}&\multicolumn{4}{c}{SSP parameters}\\
					\hline
					$\chi$&$\log_{10}(\mathcal{S}_\mathrm{B})$&\multicolumn{2}{c||}{$\log_{10}(\mathcal{S}_\mathrm{CV})$}&\multicolumn{2}{c|}{$\log_{10}\beta$}&\multicolumn{2}{c|}{$\alpha_\mathrm{IMF}$}&\multicolumn{2}{c}{$\log_{10}(\mathrm{N_{Ia}})$}\\
					(Tab.\,\ref{tab:yieldsets})&$\mu$&$\mu$&$\sigma$&$\mu$&$\sigma$&$\mu$&$\sigma$&$\mu$&$\sigma$\\
					\hline
					\multicolumn{2}{c}{}&Priors&(Tab.\,\ref{tab:priors}):&1.0&$\pm0.5$&$-2.3$&$\pm0.3$&$-2.75$&$\pm0.3$\\
					\hline
					\multicolumn{10}{c}{\textit{(a) All 29 available elements}}\\
					\hline
					C04&$-1.21$&$-1.01$&$^{+0.01}_{-0.02}$&0.77&$^{+0.30}_{-0.35}$&$-2.30$&$^{+0.11}_{-0.08}$&$-3.14$&$^{+0.16}_{-0.17}$\\
					N13&$-5.69$&$-2.01$&$^{+0.04}_{-0.03}$& 0.58 & $\pm0.29$ & $-2.32$ & $^{+0.16}_{-0.15}$ & $-2.90$ & $^{+0.18}_{-0.19}$\\
					W17&$-0.78$&$-1.01$&$^{+0.02}_{-0.03}$&$0.72$&$^{+0.32}_{-0.36}$&$-1.83$&$^{+0.18}_{-0.22}$&$-3.29$&$^{+0.15}_{-0.18}$\\
					R17&$-6.11$&$-1.62$&$^{+0.08}_{-0.04}$&$0.50$&$^{+0.26}_{-0.36}$&$-2.29$&$^{+0.19}_{-0.23}$&$-2.91$&$^{+0.21}_{-0.23}$\\
P18&0.86&$-0.90$&$^{+0.01}_{-0.03}$&$0.79$&$^{+0.27}_{-0.33}$
&$-2.13$&$^{+0.21}_{-0.20}$
&$-2.93$&$\pm0.16$\\
					\hline
					\multicolumn{10}{c}{\textit{(b) 9 most abundant elements}}\\
					\hline
					C04&$1.61$&$0.23$&$\pm0.01$&$1.01$&$^{+0.32}_{-0.35}$&$-2.45$&$^{+0.15}_{-0.11}$&$-2.89$&$^{+0.13}_{-0.12}$\\
					N13&$0.65$&$0.02$&$^{+0.02}_{-0.03}$&$0.83$&$^{+0.37}_{-0.36}$&$-2.52$&$^{+0.12}_{-0.11}$&$-2.79$&$^{+0.11}_{-0.12}$\\
					W17&$0.73$&$-0.18$&$^{+0.03}_{-0.01}$&$0.90$&$^{+0.27}_{-0.30}$&$-2.19$&$^{+0.24}_{-0.17}$&$-3.09$&$^{+0.14}_{-0.16}$\\
					R17&$-0.49$&$-0.40$&$^{+0.05}_{-0.04}$&$0.78$&$^{+0.31}_{-0.29}$
					&$-2.00$&$^{+0.23}_{-0.22}$&$-3.22$&$^{+0.21}_{-0.21}$\\
P18&$-0.01$&$-0.68$&$^{+0.06}_{-0.09}$&0.72&$^{+0.35}_{-0.36}$&$-2.22$&$^{+0.20}_{-0.19}$&$-2.84$&$^{+0.16}_{-0.18}$\\
					\hline
$\mathrm{TNG}_\mathrm{fiducial}$&$-0.89$&$-0.98$&$^{+0.01}_{-0.02}$&$0.64$&$\pm0.33$&$-2.3$&fixed&$-2.89$&fixed\\
TNG$_\mathrm{SNIa\,free}$&$-0.31$&$-0.15$ &$^{+0.02}_{-0.01}$&$0.80$ &$^{+0.29}_{-0.36}$ &$-2.3$&fixed&$-2.63$&$\pm0.13$\\
TNG$_\mathrm{IMF\,\&\,SNIa\,free}$&$1.82$&$0.29$&$ \pm0.01$&$1.13$&$\pm0.46$&$-2.68$&$^{+0.08}_{-0.09}$&$-2.87$&$^{+0.10}_{-0.09}$
					
					\label{tab:posteriors}
				\end{tabular}
			\end{center}
		\end{minipage}
	\end{table*}

\subsubsection{Scoring Metrics}\label{subsubsec:overallScores}

\begin{figure}[htbp]
\centering
\subfigure[$\beta=1 \Rightarrow $ Flat error model]{\includegraphics[width=0.4\textwidth]{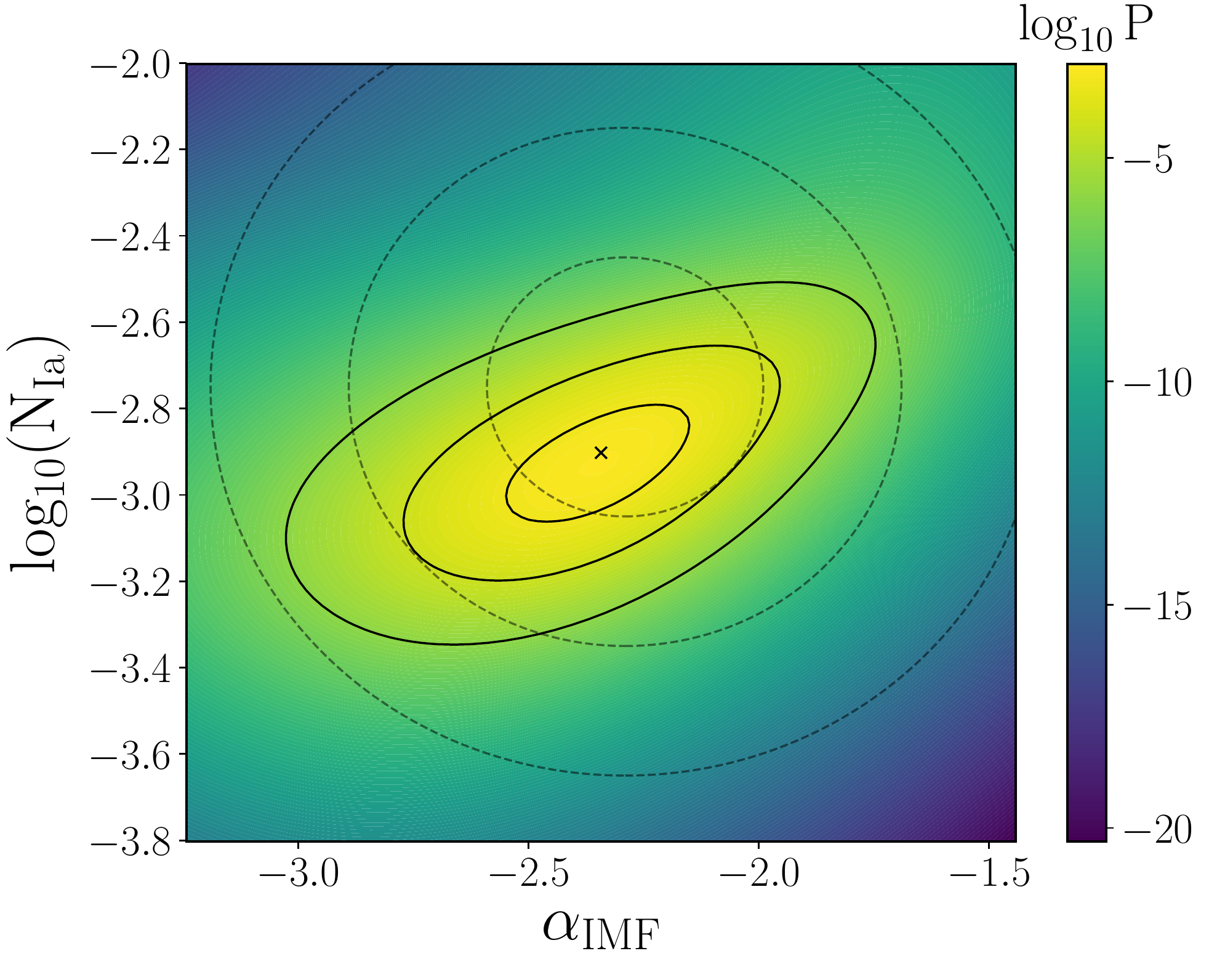}}\label{fig:post_beta1}
\subfigure[$\beta=1000 \Rightarrow$ Zero error model] {\includegraphics[width=0.4\textwidth]{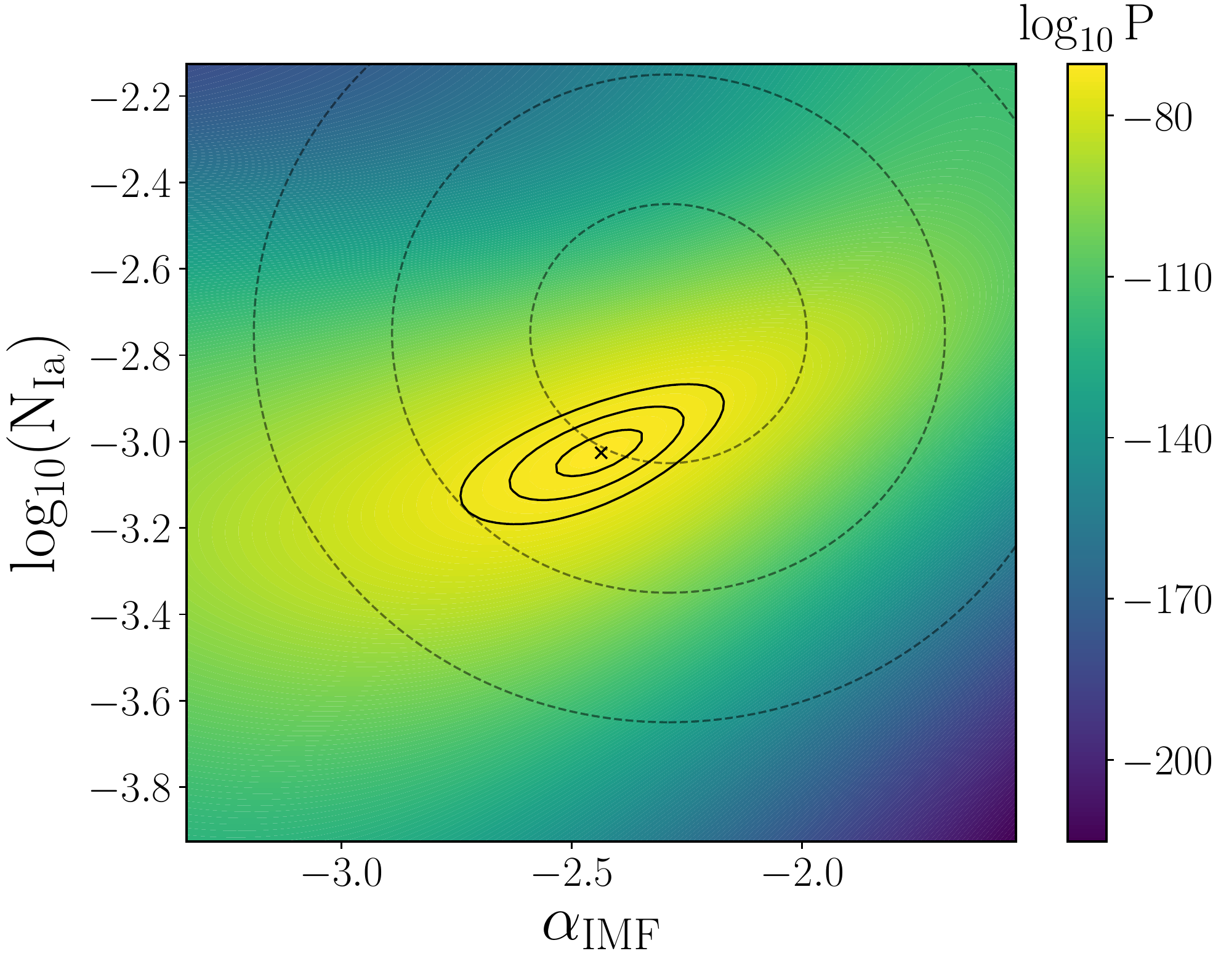}}\label{fig:post_beta1000}
\caption{$\alpha_\mathrm{IMF}-\mathrm{\log_{10}(N_{Ia})}$ plane slice of the log posterior function for the limiting values of $\beta$ with all other parameters fixed to their median posterior values. This uses N13 yields and the $n_\mathrm{el} = 28$ dataset. Solid (dashed) lines show the posterior (prior) 1, 2, 3$\sigma$ confidence intervals from the MCMC posterior PDF, and a cross gives the location of the median posterior. The scale bar demonstrates the effect of introducing the error model, and we note that contributions to the Bayesian integral (equation \ref{eq: evidence}) in the outer regions of parameter space are negligible.}\label{fig:posteriorPlot}
\end{figure}

Figure \ref{fig:posteriorPlot} shows a slice of the log posterior function in the $\alpha_\mathrm{IMF}-\mathrm{\log_{10}(N_{Ia})}$ plane for two values of $\beta$ corresponding to flat (upper panel) and approximately zero (lower panel) error models. This uses the $n_\mathrm{el} = 28$ dataset and N13 yields for illustration. Both the scale bar and confidence intervals demonstrate the strongly peaked nature of the function, and it is clear that the vast majority of the integrated posterior comes from the central regions, as previously assumed. The posterior obtained for $\beta = 1$ is many orders of magnitude greater than that for $\beta = 1000$, since we only allow for errors in the former case, giving a broader posterior. In addition, the confidence intervals shrink considerably as $\beta$ increases, since the optimal parameters become more tightly constrained.

Table \ref{tab:posteriors}\,(a) gives the results of the Bayes and LOO-CV scores of each CC-SN yield table. The uncertainty estimates on $\mathcal{S}_\mathrm{B}$ coming from the errors in the Monte Carlo integration procedure are negligible, and are thus not included. Other sources of uncertainty could, for example, be arising from \textit{Chempy} modelling assumptions or the neural network implementation. For $\mathcal{S}_\mathrm{CV}$, the uncertainty is given by the results' standard deviation from rerunning the analysis 10 times (the MCMC results will change and affect the posterior predictions). 

For this set of elements, the Bayes scores ($\mathcal{S}_\mathrm{B}$) show clearly that P18, W17 and C04 yields better reproduce proto-solar abundances, with an overall preference for P18 and a relative Bayes factor of greater than $10^4$ for either compared to N13 or R17 yields. The LOO-CV prescription ($\mathcal{S}_\mathrm{CV}$) is in accordance with this, having comparable scores for P18, W17 and C04, albeit with weaker improvements over N13 and R17, although still statistically significant. Direct comparison of the two scores is difficult due to the different metrics applied, but it is clear that the preferences are stronger for the Bayes scores. We may thus conclude that the P18, C04 and W17 models are the optimal sets in this case, with the caveat that the W17 yields predict an unreasonable top-heavy IMF (due to a large amount of black hole production). We thus take P18 yields to be the best predictor of proto-solar abundances using this set of elements.\footnote{We attribute this predominantly to the inclusion of stellar rotation in the P18 models, since rerunning the analysis using P18 yields without rotation gives much lower Bayes and LOO-CV scores ($-1.52$ and $-1.09^{+0.02}_{-0.01}$ respectively).}

\subsubsection{Posterior Element Predictions}
\begin{figure*}[htb]
\includegraphics[width=\textwidth]{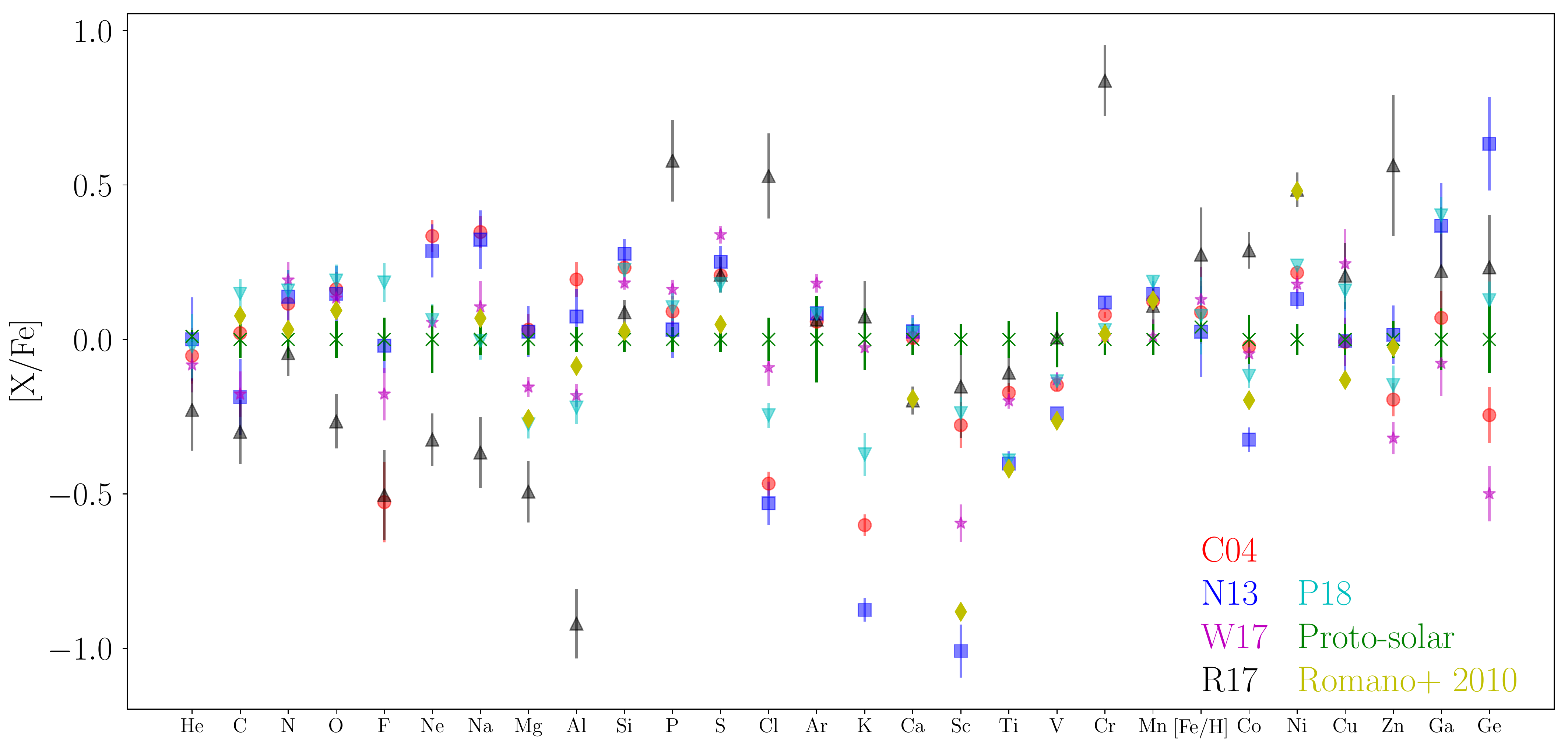}
\caption{Elemental abundance predictions obtained via the LOO-CV method excluding each element in turn, using the five CC-SN yield sets (as marked). $\beta$ is included as a free parameter for the MCMC analysis. This is done for $n_\mathrm{el} = 28$ and the proto-solar observational constraint is plotted in green. All abundances are given as [X/Fe] except for [Fe/H]. For comparison we also plot the best model of \citetalias{2010A&A...522A..32R} in yellow diamonds.}\label{fig:element_pred}
\end{figure*}

A benefit of the LOO-CV approach is that we can visualise the predictions of each element, as derived from the MCMC parameter PDF with that element excluded. Figure \ref{fig:element_pred} shows this for all available elements, comparing the five yield sets and proto-solar observations. The median values and error bars are taken from the Gaussian fitting of the PDF for each element, as described in section \ref{subsec:cross-val}. For comparison we also plot the abundances from \citetalias[fig.~23]{2010A&A...522A..32R} normalised to \cite{2009ARA&A..47..481A} solar abundances.

From the plot it is clear that there are still significant inadequacies in current yield tables, as many predicted abundances are far from their observational counterparts (missing nucleosynthetic channels could be an alternative explanation as well as oversimplifications in the GCE model). Some elements are particularly poorly predicted by certain yield sets, for example K and Cl by N13 and C04 (and to a lesser extent P18), Sc by N13 and W17, and Al and Cr by R17. This indicates that their physics is not well understood, as noted by \cite{2002A&A...388..842A,2006ApJ...653.1145K,2011ApJ...739L..57K}; \citetalias{Rybizki} amongst others; though \citet{2018MNRAS.474L...1R} overcome the deficiency of the odd-Z elements by using interacting convective O and C shells in massive stars.

The \citetalias{2010A&A...522A..32R} abundances are shown for their optimal `Model~15', using their fixed GCE model with yields from \cite{2010MNRAS.403.1413K} for AGB stars, \cite{1999ApJS..125..439I} for SN\,Ia and \cite{2006ApJ...653.1145K} for CC-SN (similar to our N13 yields, although they apply a hypernova fraction of unity for stars with $m>20\,\mathrm{M}_\odot$ and extrapolate to M$_\mathrm{CC-SN,max}=100$\,M$_\odot$), supplemented by He, C, N, O from pre-SN yields of \cite{2002A&A...390..561M,2005A&A...433.1013H,2007A&A...461..571H,2008A&A...489..685E}. Their solar abundance pattern is very similar to our N13 results except for C, N and O, which is as expected since the yields are almost identical except for the CNO elements where pre-SN yields are included which consider rotation and produce significant N in metal poor stars \citepalias[cf.][fig.3]{2010A&A...522A..32R}.

The approach of \citetalias{2010A&A...522A..32R} is completely different to that presented here as they change the yields and attempt to fit as many observational constraints as possible whilst keeping the other GCE parameters fixed (which are chosen and validated using other data). We highlight their work here to show that we both find the major limitation in reproducing the solar elemental abundance pattern to be the yield sets used and possibly missing nucleosynthetic enrichment channels.

Similarly a comparison of N13 and R17 is depicted in figure 2 of \cite{2017ApJ...836..230C} for the abundances of O, Mg, Si, Ti, Mn and Ni with respect to Fe (using M$_\mathrm{CC-SN,max}=30$\,M$_\odot$). The two yield sets show the same relative difference in abundance pattern as is visible in our figure \ref{fig:element_pred}, with R17 yields producing more Ni and Ti and less O and Mg compared to N13 in both studies.

For all yield sets we see an underproduction of Sc and Ti. One possible reason might be that neutron star winds (not included in the yield tables) contribute significantly to those elements \citep{2005ApJ...623..325P}. \cite{2006ApJ...640..891Y} argue that transitioning from 1D to 3D models will increase Ti and Ni yields. However Ni is already overproduced by all CC-SN yield sets, though we also expect significant contributions (40-70\%) of Ni by SN~Ia \citepalias[fig.13]{Rybizki}. In addition, we note a consistent over-abundance of Si and S for all CC-SN tables. \citet[fig.\,8]{Fryer} clearly demonstrate that this might be remedied by decreasing the CC-SN explosion energies.

Despite the inclusion of a variable error model, both our statistical methods are significantly affected by poorly reproduced elements, and exclusion of certain elements, such as Sc, could thus drastically change and reorder the overall scores. This indicates the necessity of matching the elemental subset for which the score is calculated to that of the simulation.

\subsection{Subset of most abundant elements}\label{subsec:IllustrisResult}
The above analysis can similarly be performed for any subset of elements, for example the 9 most abundant, which are usually tracked in hydrodynamical simulations. These are the most well understood elements, thus we would expect our CC-SN yield tables to perform better in this context.
 
\subsubsection{Posterior Parameter Distributions}\label{subsubsec:MCMCIllus}
Tables \ref{tab:posteriors}\,(b) and \ref{tab:ism_parameters}\,(b) give the relevant MCMC output for the five yield tables, and we see that in almost all cases the posteriors are far greater than those previously obtained, with much less distinction between yield sets. This is a direct result of the removal of the most poorly predicted elements from the analysis. Similarly, the median values of $\beta$ are greater for most yield sets, indicating a preference for smaller model errors, suggesting a better fit to the data. The 1$\sigma$ confidence intervals of $\alpha_\mathrm{IMF}$ and $\log_{10}(\mathrm{N_{Ia}})$ for all yield tables can be seen to overlap with the respective intervals from the $n_\mathrm{el}=28$ optimisation (i.e. table \ref{tab:posteriors}\,(a)). 

For C04, N13, W17 and P18, less CC-SN and more SN\,Ia are favoured in order to reproduce proto-solar abundances (compared to the $n_\mathrm{el}=28$ run, considering $\alpha_\mathrm{IMF}$ and $\log_{10}(\mathrm{N_{Ia}})$). This is most likely because formerly underproduced elements (e.g. Sc, K or Cl as visible from figure \ref{fig:element_pred}) are now excluded and illustrates the bias that can be introduced by element choice (in combination with wrongly predicted elements). In contrast, the R17 yields produce more CC-SN and less SN\,Ia when using the reduced $n_\mathrm{el}=8$ subset. From the pattern of figure \ref{fig:element_pred}, it is clear that R17 underproduces the light $\alpha$-elements and overproduces some heavier elements (e.g. P, Cl, Cr) leading to this opposite effect. We also note that R17 produces black holes for all stars with  $m>25$~$\mathrm{M}_\odot$, giving less overall enrichment resulting in the need for the IMF to become top-heavy (see \citet{Sukhbold} for a in-depth discussion of the black-hole explosion `landscape'). 

For P18 yields, we note that the maximum posterior is in fact reduced, indicating that the 29-element score is less affected by poorly reproduced elements. Comparison of figures \ref{fig:element_pred} and \ref{fig:illustris_element_pred} show that this is due to the significant bias from the severely under-produced Mg in the 9-element case. This is reduced for $n_\mathrm{el}=28$ due to an optimal parameter set giving greater Mg production, and, for the scoring metrics, the effect of the geometric averaging over elements.

If we omit the pathological cases of W17 and R17 (due to their high `failed SN' fraction) then $\alpha_\mathrm{IMF}$ ($\log_{10}(\mathrm{N_{Ia}})$) is constrained to around -2.35 (-2.9) by C04, N13 and P18. The IMF high-mass slope, inferred here for the Milky Way from chemical evolution constraints, is slightly steeper than the canonical value, in agreement with recent inferences using star counts  in M31 \citep{2015ApJ...806..198W}.\footnote{We note that it is difficult to infer the IMF high-mass slope from star counts from within the Milky Way due to incomplete knowledge of the most massive stars \citep[e.g.][]{2015MNRAS.447.3880R,2017MNRAS.464.1738D}.}

Studies by both \citetalias{2010A&A...522A..32R} and \citetalias{2015MNRAS.451.3693M} favour a slightly bottom-heavy \cite{1993MNRAS.262..545K} IMF with a high-mass slope of $-2.7$ which results in about 2.5 times less CC-SN explosions compared to the \citetalias{2003PASP..115..763C} IMF with $\alpha_\mathrm{IMF}=-2.3$, which we more or less recover \citepalias[cf.][tab.\,5]{2015MNRAS.451.3693M}. At the same time, we are utilising a smaller CC-SN mass range with M$_\mathrm{CC-SN,max}=40\,\mathrm{M}_\odot$ which reduces the number of CC-SN by 27\% \citep[cf.][tab.1]{2005A&A...430..491R} and excludes the enrichment from super-massive stars which usually relies on extrapolation. \citetalias{2010A&A...522A..32R} does this up to 100\,M$_\odot$ whereas \citetalias{2015MNRAS.451.3693M} simply integrates up to the respective highest mass grid point. In addition, the inferred IMF also depends on the yield sets and observational constraints applied, which differ for all three works. 

As another example, \cite{Suzuki} explicitly address the issue of which mass ranges explode as CC-SN and where explosions fail using GCE arguments with C04 yields together with solar O and Fe observations. They conclude that M$_\mathrm{CC-SN,max} = 100$~$\mathrm{M_\odot}$ best reproduces observations but their analysis suffers from yield table extrapolation and fixed GCE parameters. When allowing for the latter (as here) those abundances are well recovered using C04 yields and M$_\mathrm{CC-SN,max}=40$\,$\mathrm{M_\odot}$ (cf.~figure \ref{fig:illustris_element_pred}). This illustrates that keeping GCE parameters fixed will inevitably bias the inference.

\subsubsection{Scoring Metrics}\label{subsubsec:OverallIllustris}
The Bayes scores are several orders of magnitude greater than for $n_\mathrm{el} = 28$ in most cases as a result of the exclusion of poorly reproduced elements. In accordance, $\beta$, $\mathcal{S}_\mathrm{B}$ and $\mathcal{S}_\mathrm{CV}$ are largest for C04 with this choice of elements, albeit with a value of $\log_{10}(\mathcal{S}_\mathrm{CV})=0.23$ which is still not compatible with the self-observation score (obtained by setting the model `data' equal to a Gaussian realisation of the proto-solar observations) which has $\log_{10}(\mathcal{S}_\mathrm{CV})=0.64^{+0.04}_{-0.06}$. In $\mathcal{S}_\mathrm{B}$ we note a relative preference of $\approx10$ for C04 over N13 and W17, which are themselves favoured over R17 and P18 by a significant factor. The LOO-CV scores are less decisive with relative scores of around 1.6 (4) between the C04 and N13 (R17) yields for example, meaning that detailed conclusions are more difficult. Both metrics are in agreement however that the R17 and P18 yields are the worst predictor of proto-solar abundances for the CC-SN yields and element set tested here. As for the maximum posteriors, we note that, in the case of P18, $\mathcal{S}_\mathrm{B}$ is lower for the 9-element subset, again as a consequence of the greater influence of the Mg-induced bias in this case. For the LOO-CV score, we report no significant change, due to the weaker dependence of this score on poorly reproduced elements.

The ranking varies for the two scores and also for the different elemental subsets, although overall C04 seems to score well in both, whereas R17 tends to come last. \citetalias{2015MNRAS.451.3693M} also prefers C04 CC-SN yields even though their elemental subset (C, N, O, Fe and [$\alpha$/Fe]) is different to ours and other observational constraints like the metallicity distribution function (MDF) and star formation history are included. At the same time, they find that the AGB yields are less important (despite half their elements being strongly affected by AGB feedback) and that a \cite{1993MNRAS.262..545K} IMF is preferred. They also face the problem with how to factor in results from each of their $\chi^2$ metrics, hence a second overall score is reported with the MDF being left out, resulting in C04 and N13 scoring best together with the \cite{2010MNRAS.403.1413K} AGB yields. \citetalias{2010A&A...522A..32R} also favours N13 CC-SN yields over the classical \cite{1995ApJS..101..181W} values (C04 yields were not included in their study).

\subsubsection{Posterior Element Predictions}
\begin{figure}[bht]
	\includegraphics[width=0.5\textwidth]{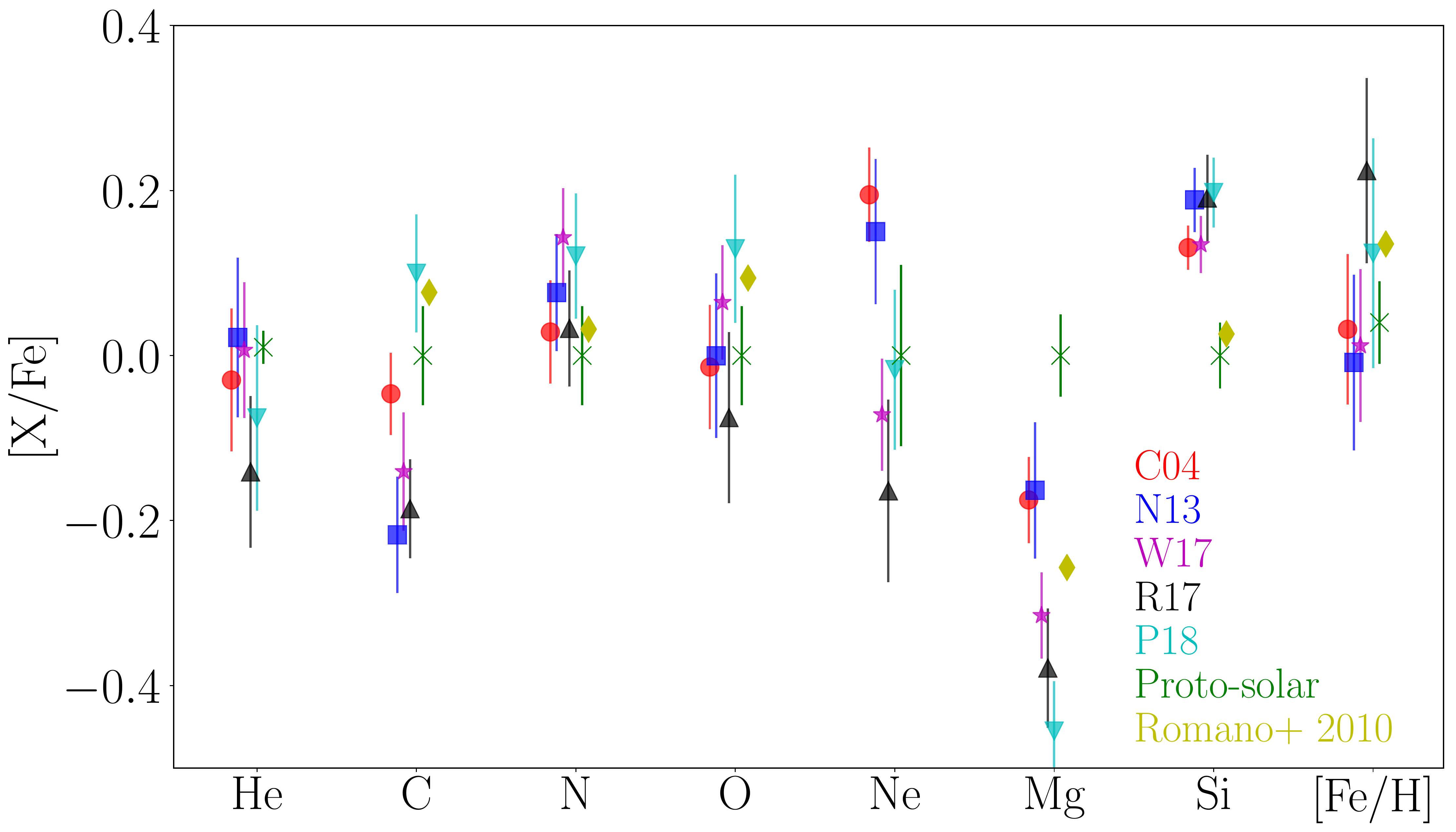}
	\caption{As for figure \ref{fig:element_pred}, but for the reduced elemental subset of 9 most abundant elements and with a lateral displacement of data-points to increase discernibility.}\label{fig:illustris_element_pred}
\end{figure}

Figure \ref{fig:illustris_element_pred} plots the LOO-CV predictions for each element, showing far better agreement between observations and predictions than in figure \ref{fig:element_pred}, as expected. The majority of the predictions are consistent with observations at a $2\sigma$ level, although there are still significant discrepancies for Mg using W17, R17 \& P18 yields, and a significant overproduction of Fe by R17.

The yield set predictions for Mg and Si are broadly self-consistent, yet significantly different to the proto-solar data, with Mg (Si) always under- (over-)produced. This could be mitigated in CC-SN modelling by decreasing the explosion energy of the piston \citep[fig.8]{2010ApJ...724..341H,Fryer}, or could be indicating errors in the solar abundance determination though that seems more unlikely \citep{2017MNRAS.464..264A}.

\begin{figure}[bht]
	\includegraphics[width=0.5\textwidth]{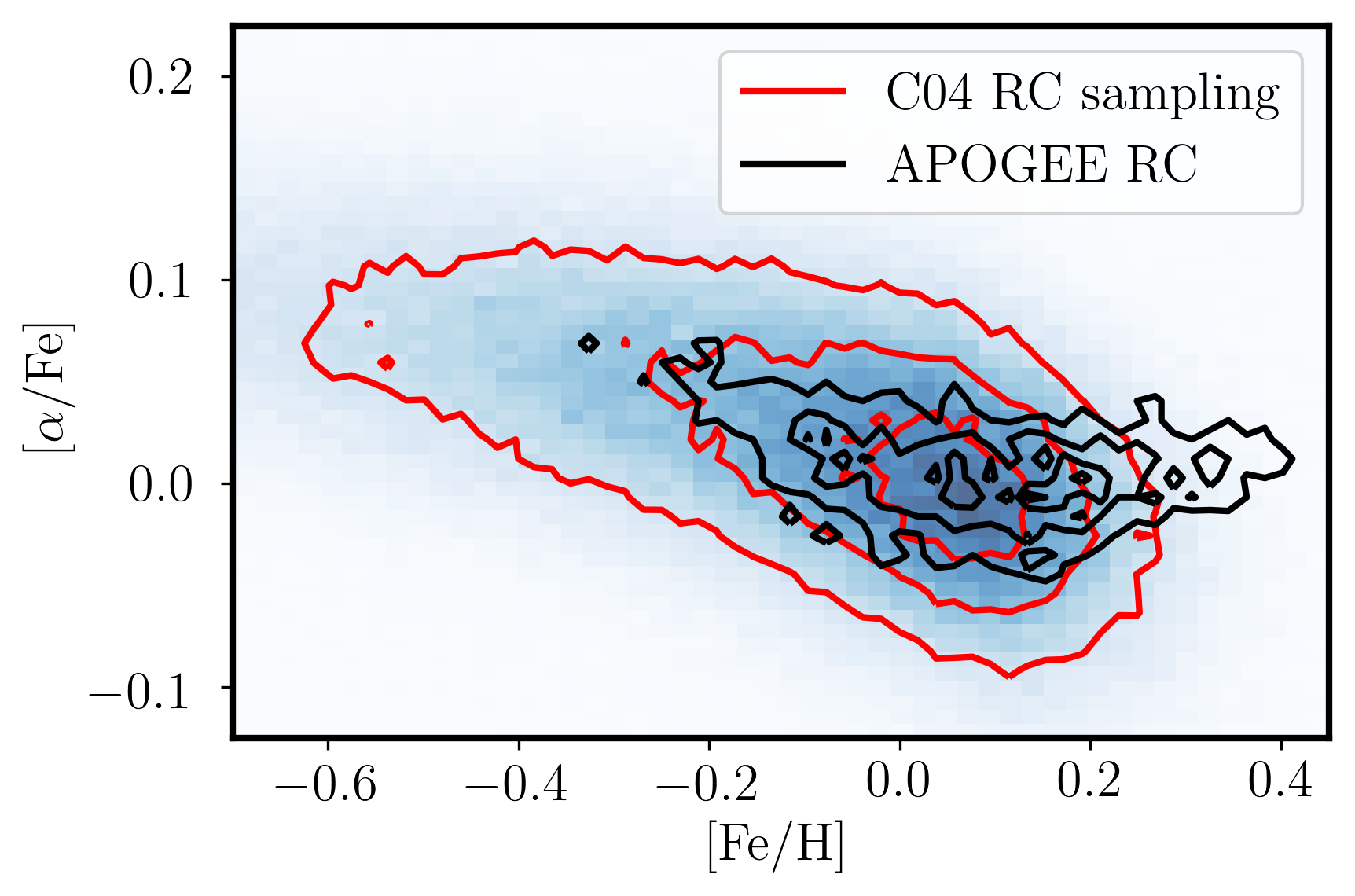}
	\caption{Alpha enhancement over metallicity, comparing a \textit{Chempy} model with C04 yields, $n_\mathrm{el}=8$ and 250 draws from the optimised posterior distribution (blue density map and red contours) with APOGEE red clump stars (black contours). This uses the same selection of APOGEE RC stars as in figure\,\ref{fig:TNGvsAPOGEE} except that the contours are here added to the linearly binned 2D distribution instead of the logarithm. A summary of how \textit{Chempy} predictions were transformed into red clump mock observations is given in the text.}\label{fig:C04vsapogee}
\end{figure}

  In figure \ref{fig:C04vsapogee} we show predictions of our modelling in comparison to APOGEE red clump (RC) stars \citep{2014ApJ...790..127B}. We select 4832 such objects from APOGEE DR14 \citep{Abolfathi} subject to the constraints $5\,\mathrm{kpc}<R_\mathrm{Gal}<9\,\mathrm{kpc}$, $|z|<1$ and with the APOGEE flags \texttt{ANDFLAG}, \texttt{ASPCAPFLAG}, \texttt{EXTRATARG} being zero. For the \textit{Chempy} model we show our C04 results with $n_\mathrm{el}=8$. The distribution is sampled using 250 randomly drawn evolutionary tracks from the MCMC posterior. For each track the respective SFR was taken into account together with the red clump age distribution. An observational error of 0.05\,dex in [Fe/H] and 0.03\,dex in [$\alpha$/Fe] was added to simulate real data. Further details on this procedure are given in \citet[sec.\,3]{2016AN....337..880J}. Similar to \citetalias[fig.\,11]{Rybizki}, we see that the MDF of the APOGEE data does not possess the long tail to lower metallicities which is seen in the \textit{Chempy} mock observations; instead they produce a greater number of super-solar metallicity stars. In addition, the highest density of both model and data are at the origin, i.e. solar values of [Fe/H] and [$\alpha$/Fe], as expected since the \textit{Chempy} parameters are optimised to reproduce proto-solar abundances. We caution that this compares a selection of red clump stars to predictions of a \textit{Chempy} model that was optimised to reproduce only 8 proto-solar abundances, thus the spatial selection in the data does not represent our Solar one-zone model (a proxy of which is impossible to find), nor do the individual elements fit equally well. For these reasons, the APOGEE data has not been used as constraining data for \textit{Chempy} in this work.
\subsection{Testing inferred SSP parameters in hydrodynamical simulations}\label{subsec:TNGYields}

It is not certain \textit{a priori} that our predicted best-fit SSP parameters, obtained by marginalising over \textit{Chempy} ISM parameters, will be useful for a hydrodynamical simulation with a completely different ISM physics representation, since specific model assumptions can significantly bias the inferred parameters, as shown in \cite{2017ApJ...835..128C}. Even if the ISM model were similar, \textit{Chempy} uses constant parameter values over the whole period of galactic evolution, some of which (e.g. the star formation efficiency) will vary over time and space in a hydrodynamical simulation.

\subsubsection{Optimising parameters for a specific yield set}
\label{sec:specific_yield}
To test this, we apply our method to predict two SSP parameters that we subsequently insert into a zoom-in simulation of a Milky Way-like galaxy using the cosmological magneto-hydrodynamical code {\sc AREPO}. The fiducial model is identical to the simulation of ``Halo 16" (level 5) in \citet[][hereafter \citetalias{2017arXiv171004222G}]{2017arXiv171004222G}, which uses initial conditions from the Auriga project  \citep{2017MNRAS.467..179G} and has been shown to match a variety of observational constraints from the Milky Way. The yield set is identical to the one introduced in the IllustrisTNG simulations \citep[][tab.\,2]{Pillepich}, henceforth referred to as TNG yields. These are implemented into \textit{Chempy}, with $\mathrm{M_{CC-SN,max}}$ and the SN\,Ia delay time modified to match the simulation values of 100\,M$_\odot$ and 40\,Myr, respectively\footnote{Results of \textit{Chempy} runs with $\mathrm{M_{CC-SN,max}}=40$ M$_\odot$ can be inspected in appendix \ref{appenB}, table \ref{tab:ism_parameters}\,(d).}. This ensures that our IMF and SN\,Ia model (and yield set) are identical to those of \citetalias{2017arXiv171004222G}, hence the SSP enrichment over time is the same for both codes if the same SSP parameters are used. The fiducial values in \citetalias{2017arXiv171004222G} are $\log_{10}(\mathrm{N}_\mathrm{Ia})=-2.89$ and $\alpha_\mathrm{IMF}=-2.3$ (using a \citetalias{2003PASP..115..763C} IMF).
 
Three \textit{Chempy} analyses were performed using $n_\mathrm{el}=8$:
 \begin{enumerate}
 	\item Leaving all 5 \textit{Chempy} parameters and $\beta$ free, using TNG yields with the appropriate M$_\mathrm{CC-SN,max}$ and SN\,Ia delay time values ($\mathrm{TNG_{IMF\,\&\,SNIa\, free}}$)
 	\item As (1) but fixing $\alpha_\mathrm{IMF}$ to -2.3 ($\mathrm{TNG_{SNIa\,free}}$)
 	\item As (2) but also fixing $\log_{10}(\mathrm{N_{Ia}})$ to -2.89 ($\mathrm{TNG_\mathrm{fiducial}}$).
 \end{enumerate}



A graphical summary of the resulting chemical abundance tracks drawn from the posterior of this inference can be inspected in figure \ref{fig:TNGvsAPOGEE}. The resulting best-fit parameters and scores are summarised at the end of table \ref{tab:posteriors}. As expected, the scores increase with every additional free parameter and the effect of altering the IMF is found to be stronger than that of the SN\,Ia normalisation. The Bayes score increases by almost $10^3$ when allowing both SSP parameters to vary freely compared to the fiducial case where they are fixed. We note that it also supersedes the score found for our other yield sets even though the TNG yield set is very similar to our N13 yield set. The reason is most likely that the TNG yields have additional grid points at 9 and 12 M$_\odot$ from the \cite{1998A&A...334..505P} yields which allows for a higher flexibility as the N13 yields must be extrapolated from 13\,M$_\odot$ down to 8\,M$_\odot$. In the case with only N$_\mathrm{Ia}$ free, the number of SN\,Ia per M$_\odot$ almost doubles, but when the IMF is also allowed to vary, N$_\mathrm{Ia}$ becomes close to the fiducial value and $\alpha_\mathrm{IMF}$ decreases to -2.68 resulting in a strongly bottom-heavy IMF. Both parameters are very well confined with uncertainties of $\pm 0.1$.

\begin{figure}[thb]
	\includegraphics[width=0.5\textwidth]{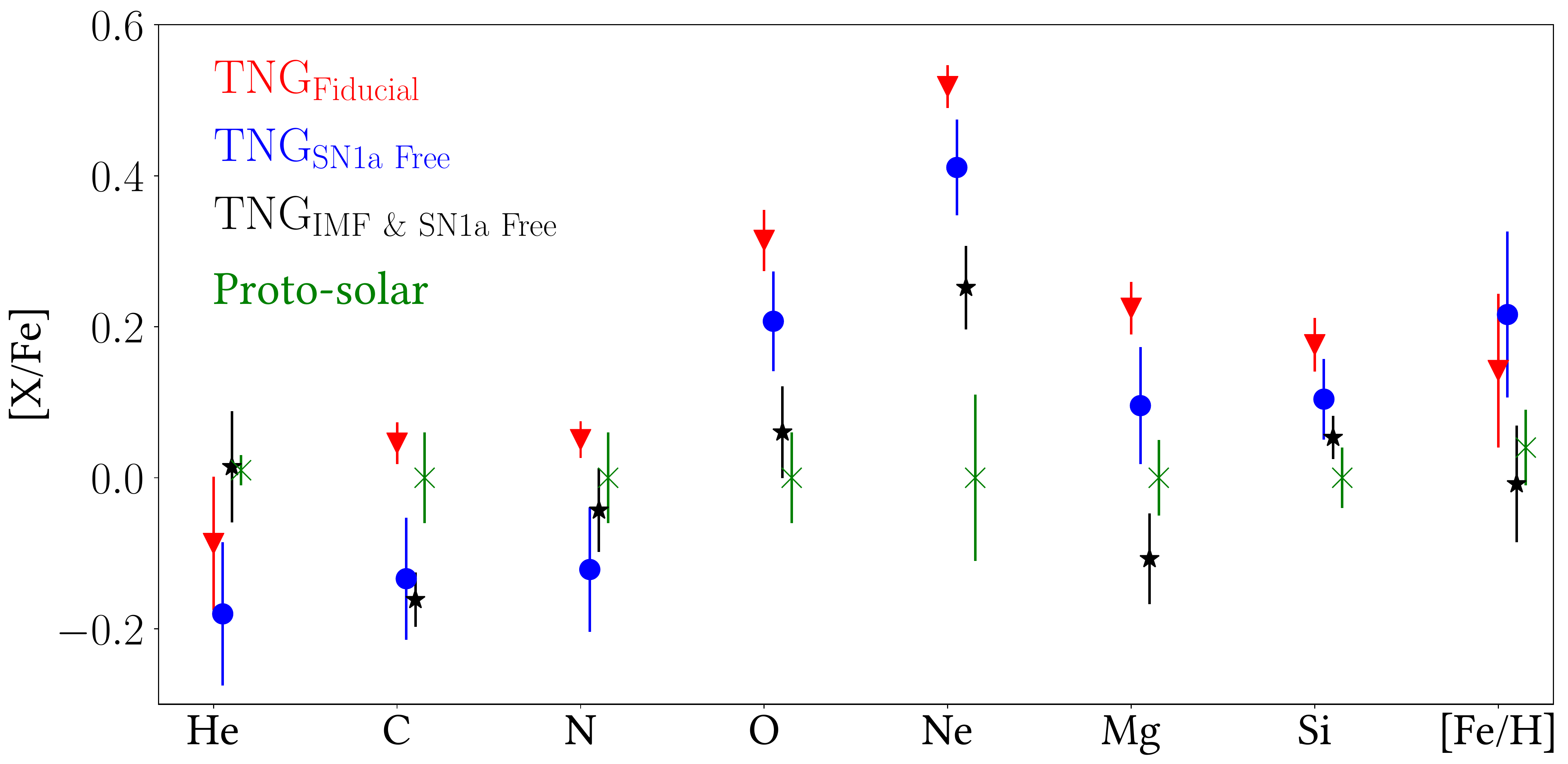}
	\caption{As figure\,\ref{fig:illustris_element_pred} but now using the TNG yield set of \cite{Pillepich} and an increased $\mathrm{M_{CC-SN,max}}=100$~M$_\odot$ to match parameters of \citetalias{2017arXiv171004222G}. The LOO-CV predictions of three runs are shown: (1) with \textit{Chempy} parameters and $\beta$ free (black stars); (2) like (1) but fixing $\alpha_\mathrm{IMF}$ to the fiducial value of -2.3 (blue circles); (3) like (2) but also fixing N$_\mathrm{Ia}$ to its fiducial value of $1.3\times10^{-3}$\,M$_\odot^{-1}$ (red triangles).}\label{fig:TNGYields} 
\end{figure}


The predicted proto-solar abundances from \textit{Chempy} for these runs are shown in figure \ref{fig:TNGYields} and we note a strongly $\alpha$-enhanced pattern when fiducial SSP values are used. By increasing the number of SN\,Ia, the Fe yield increases which, in turn, decreases the [X/Fe] values giving a better fit to the data despite only minimally affecting the overall pattern. The relative abundances change more strongly when allowing the IMF to vary, giving the fit of predictions to observations.

\begin{figure*}[thb]
	\includegraphics[width=\textwidth]{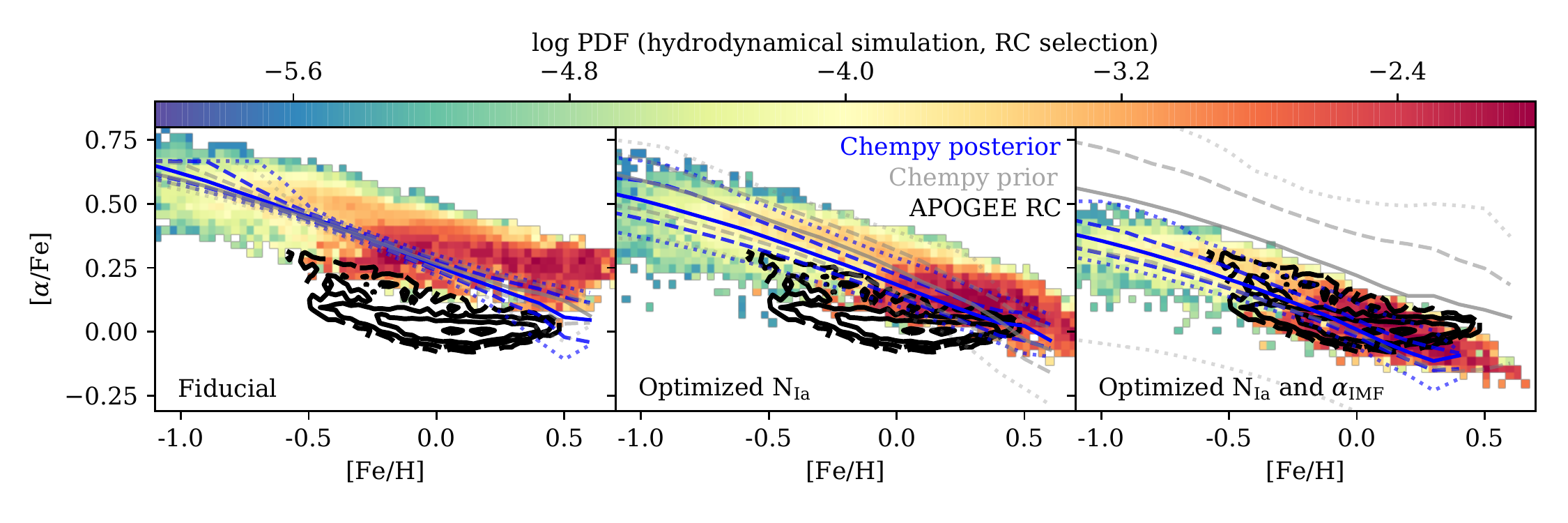}
	\caption{Distribution of stellar particles in the Halo 16 simulation at $z=0$ in the [$\alpha$/Fe] vs. [Fe/H] plane for the fiducial simulation (left), the simulation using the optimised value of $\mathrm{N_{Ia}}$ (centre), and that with the optimal values of both $\mathrm{N_{Ia}}$ and $\alpha_\mathrm{IMF}$ (right). The black contours denote the logarithmic density isocurves of APOGEE red clump stars \citep{2014ApJ...790..127B} from DR14 \citep{Abolfathi}, using the same selection as in figure \ref{fig:C04vsapogee}. Both the spatial and red clump age selections are also applied to the hydrodynamical simulation, with this figure following the same analysis as \citetalias[][fig.\,6]{2017arXiv171004222G}. To this we add blue (grey) lines showing the median abundance tracks from the respective \textit{Chempy} posterior (prior)
with the dashed (dotted) lines enclosing the 1$\sigma$ (2$\sigma$) deviations of all tracks}\label{fig:TNGvsAPOGEE}
\end{figure*}

\subsubsection{Comparing hydrodynamical simulation predictions to observational data}
Next, these results are tested by re-running the hydrodynamical simulation after inserting the derived SSP parameters. 
In figure \ref{fig:TNGvsAPOGEE}, the predicted abundance distribution in the [$\alpha$/Fe] vs. [Fe/H] plane is shown for the hydrodynamical simulations with each set of SSP parameters, with the left panel showing the fiducial case \citepalias[cf.][fig.\,6]{2017arXiv171004222G}. The middle panel displays the simulation using the optimised parameter $\log_{10}(\mathrm{N}_\mathrm{Ia})=-2.63$, while the right panel uses the parameters $\log_{10}(\mathrm{N}_\mathrm{Ia})=-2.87$ and $\alpha_\mathrm{IMF}=-2.68$, both left free during the \textit{Chempy} analyses. Here, $\alpha$ is defined as the mass weighted average of O, Mg and Si. The abundance distributions are compared to those of the APOGEE red clump stars \citep{2014ApJ...790..127B,Ness,Abolfathi} plotted in black contours. Stars from the simulation are weighted according to the red clump age distribution \citep[][eq.\,11]{2014ApJ...790..127B} and selected according to the APOGEE spatial selection of $5\,\mathrm{kpc} <R_\mathrm{Gal} < 9\,\mathrm{kpc}$ and $|z|<1\,\mathrm{kpc}$. This data was not used during the \textit{Chempy} optimisation and is depicted to illustrate the typical variation of Milky Way disc stars. Additionally, we show in blue (grey) the median from 250 abundance tracks randomly drawn from the posterior (prior) of the respective \textit{Chempy} inference specified in section\,\ref{sec:specific_yield}. The dashed (dotted) lines enclose the 1$\sigma$ (2$\sigma$) contours of these tracks.

Whereas no Sun-like star is produced by the simulation with the fiducial SSP parameters (since the abundance distribution is far too $\alpha$-enhanced), there are stars with [Fe/H] = [$\alpha$/Fe] = 0 for the run using the $\log_{10}(\mathrm{N_{Ia}})$ prediction, similar to the findings in \citet[fig.\,4]{Naiman} where $\mathrm{N_{Ia}}$ is increased by a factor of 3.5. However, $\alpha$-enhancement is only decreased by a small fraction, still being 0.2\,dex over-abundant with respect to the APOGEE stars. Considering the simulation using both optimised SSP parameters, the resulting abundance distribution of the Milky Way-like hydrodynamical simulation matches the APOGEE data very closely, both in slope and [$\alpha$/Fe]-scatter of the distribution. Some discrepancy remains, with a few more stars in the metal-poor regime and a metal-rich tail not found in the observational data. The latter may arise from too efficient radial migration in the simulation.

Since no additional parameters were changed for these simulations, we note that many simulation predictions of the fiducial run are likely altered when modified SSP parameters are used. Most notably, the feedback parameters were not adjusted, and the new $\mathrm{N_{Ia}}$ and $\mathrm{\alpha_{IMF}}$ values possibly has significant impact on the resulting properties of the galaxy. Seeing that our goal here is to show how \textit{Chempy} can be used to optimize the $\alpha$-abundance predictions, such investigation is beyond the scope of this paper.

\subsubsection{Why can our best-fit parameters improve hydrodynamical simulations?}
As can be seen from figure \ref{fig:TNGvsAPOGEE}, the \textit{Chempy} median posterior (solid blue line) is close to the running median of the hydrodynamical simulation for tested SSP parameter sets. Notably, the variances of the two distributions are similar when both IMF parameters are allowed to vary (right panel), but the \textit{Chempy} track is far narrower than that of the hydrodynamical simulation in the fiducial case with SSP parameters held constant, implying that variation of the ISM parameters causes only a minor spread in the evolutionary tracks. As noted previously, in the fiducial model we see that the posterior ISM parameters mostly reproduce the prior, thus there is negligible difference between the \textit{Chempy} prior and posterior tracks in this case. For the right panel, it is clear that the posterior distributions (and observational data) lie far from the prior evolutionary tracks; a consequence of the steeper IMF found preferred for this scenario. Despite the differing distribution widths, the clear correlation between the two simulations implies that our simple GCE model \textit{Chempy} can produce a satisfactory reproduction of the evolutionary tracks of far more complex simulations. This may be rationalised by noting that, although the ISM physics of the hydrodynamical simulation has a completely different parametrisation and includes more effects than our simple GCE model, the hyperparameters are still tuned to reproduce global ISM constraints including the Kennicut-Schmidt law \citep[fig.\,3]{2003MNRAS.339..289S} and the cosmic star formation history \citep[fig.\,6]{2013MNRAS.436.3031V}, which were also incorporated into the \textit{Chempy} ISM parametrisation and priors. Because the SSP enrichment engines are the same for both models, the optimised SSP parameters coming from a \textit{Chempy} MCMC inference are seen to be useful predictors for the chemical abundance tracks of hydrodynamical simulations. The increased ISM complexity simply adds to the variance of the abundance distribution, whilst the median track is conserved.


This outcome could be biased because only a single galaxy was considered and it is known that the chemical enrichment of galaxies of a certain mass depends on their evolutionary environment \citep[][fig.\,5, fig.\,11 respectively]{2017ApJ...845..136B,2013MNRAS.436.3031V}. However, we expect that the `environmental' abundance prediction noise at that galaxy mass to be lower than the [$\alpha$/Fe] discrepancy of 0.2\,dex found here. At the same time, we are optimising for 8 abundances (as in figure \ref{fig:TNGYields}) but only consider the $\alpha$-enhancement in the APOGEE data, so could have optimised for less elements. Similarly, the previously used proto-solar observational data could be supplemented with more data points (or the whole distribution) from APOGEE to more stringently optimise the \textit{Chempy} parameters, allowing the implementation of constraints from sub- and super-solar metallicity. Even in this simple form however, we note a marked improved in the hydrodynamical simulation abundance predictions using our optimised SSP parameters.


\section{Summary and Conclusions} \label{sec:conc}
In this paper we have constructed a scoring system for nucleosynthetic yield tables and tested it on five different CC-SN yields using a simple observational dataset; the proto-solar abundances. The software developed can be utilised to make informed decisions regarding yield table choice for simulations and to produce optimal Galactic and stellar parameters for any subset of traced elements. 

Our simulation framework has been the flexible GCE modelling software \textit{Chempy} \citep{Rybizki} with the addition of an error model to account for simulation inadequacies and yield table errors. Two scoring metrics were considered, based on Bayesian evidences and leave-one-out cross validation (LOO-CV), giving overall scores that, unlike other approaches, marginalise over the main GCE parameters; IMF slope, SN\,Ia normalisation, SFR, star formation efficiency and outflow fraction. In addition, the error parameter is left free with lower model error implying better yield table predictions. We show that the resulting abundance distributions (in the [$\alpha$/Fe]$-$[Fe/H] plane) of the \textit{Chempy} simulation (with optimised parameters) are comparable with observational data, despite being constrained only by proto-solar data.

The yield table ranking of both metrics depends on the choice of elements, thus we have tested two different sets; (a) using all available elements, and (b) the subset of 9 most abundant elements (as commonly tracked by hydrodynamical simulations). The two scores yield qualitatively similar results for most yield sets and favour \cite{2018MNRAS.tmp..302P} (P18, see also \citealt{2018arXiv180509640L}) and \cite{2004ApJ...608..405C} (C04) yields for cases (a) and (b) respectively. C04 and West \& Heger (2017, in prep.) yields also score well for case (a), but the latter requires an unrealistically top-heavy IMF (due to their model's inclusion of many failed supernovae). For our two CC-SN yield tables which do not include failed supernovae; C04 and \cite{Nomoto} (N13), we infer their mutual encompassing 1$\sigma$ range to be $-2.30>\alpha_\mathrm{IMF}>-2.63$ and $-2.68>\log_{10}(\mathrm{N}_\mathrm{Ia})>-3.01$ when using the smaller (more reliable) elemental subset. CC-SN models which include rotation (the P18 yields) substantially improve the yield table predictions; poor results for the 9-element case can be attributed specifically to the severe under-production of Mg in this model.
 
 
 
We find the Bayesian metric to be more discriminative and computationally cheaper, whereas the cross validation has the advantage of being a useful tool for identifying poorly predicted elements. 
 In case (a) we find a consistent over- (under-)production of Si, S and Ni (Sc and Ti) regardless of the tested yield table. For Sc and Ti, missing neutron star wind enrichment could potentially explain this \citep{2005ApJ...623..325P}. In case (b) we find that all CC-SN yield tables overpredict Si and underpredict Mg which could be remedied by reducing the explosion energy \citep[fig.\,8]{2010ApJ...724..341H,Fryer}. This also applies to S from case (a).  

The utility of the code in predicting optimal simulation parameters is shown via application to a hydrodynamic zoom-in simulation \citep{2017arXiv171004222G} using fixed yield set \citep[tab.\,2]{Pillepich} both with and without the constraint of fixing the IMF. The resulting abundance distribution of this simulation, which is heuristically similar to the \textit{Chempy} predictions, is significantly improved when both of our best-fit SSP parameters are used, as validated by comparison to an independent data set (see figure \ref{fig:TNGvsAPOGEE}). The improvement by only optimising the number of SN\,Ia is much less and the IMF seems to have a stronger impact on the abundance prediction. The reason why a simple GCE optimisation of SSP parameters can improve the abundance patterns produced by hydrodynamical simulations, despite strongly differing ISM physics implementations, is because both types of simulation try to reproduce average ISM relations \citep{2003MNRAS.339..289S,2013MNRAS.436.3031V}.   

In addition, our analysis may be used to test the improvement when modelling extra nucleosynthetic channels, for example neutron star mergers \citep{2017ApJ...836..230C} or super-AGB stars \citep{2010A&A...512A..10S,2015ASPC..497..247D}. Extending the observational constraints to include more stars or other GCE data (cf. \citetalias{2015MNRAS.451.3693M}) is also straightforward.

The code has been made publicly available,\footnote{\url{http://github.com/oliverphilcox/ChempyScoring}} including a comprehensive online tutorial and a method to produce SSP enrichment tables similar to \cite{2017arXiv171109172R}. 

\acknowledgments
\begin{footnotesize}
We would like to thank Kareem El-Badry, Coryn Bailer-Jones, Benoit C\^{o}t\'{e}, Chris Fryer, Brad Gibson, Robert Grand, David Hogg, Samuel Jones, Annalisa Pillepich and Hans-Walter Rix for fruitful discussions regarding this work, as well as Alexander Heger, Jill Naiman, Nikos Prantzos, Christian Ritter and Christopher West for providing us with yield tables and Donatella Romano providing us with her solar abundance predictions. In addition, we would like to thank the anonymous referee for his feedback improving the clarity of this report. JR acknowledges funding from the
European Research Council under the European Union’s Seventh Framework
Programme (FP 7) ERC Advanced Grant Agreement No. [321035].
TAG acknowledges funding through the Collaborative Research Centre SFB 881 “The Milky Way System” (subproject A1) of the German Research Foundation (DFG). The hydrodynamical simulations were
carried out on the High Performance Computing cluster {\sc Draco} at  the  Max  Planck  Computing  and Data Facility (MPCDF) in Garching operated by the Max Planck Society (MPG).
\end{footnotesize}
%





\bibliographystyle{aasjournal} 
\bibliography{adslib,otherlib,newbib} 

\appendix
\section{Testing the Neural Network}\label{appenA}
\begin{figure}[htb]
\caption{Corner plot of the median network error in the 5-dimensional space of \textit{Chempy} parameters. Each subplot shows a two-dimensional cut through parameter space with the training dataset represented by black crosses, and L1 errors (in dex) are calculated for an independent random test dataset (coloured dots). This uses all 29 elements and the N13 yield set. The diagonal plots show histograms of the uniformly distributed validation datasets, with values of $\theta_\mathrm{prior}$ ($\sigma_\mathrm{prior}$) are shown by solid (dashed) vertical lines. The errors (shown by colours) are tiny in central regions compared to the minimum observational error of 0.02 dex.}
\includegraphics[width=\linewidth]{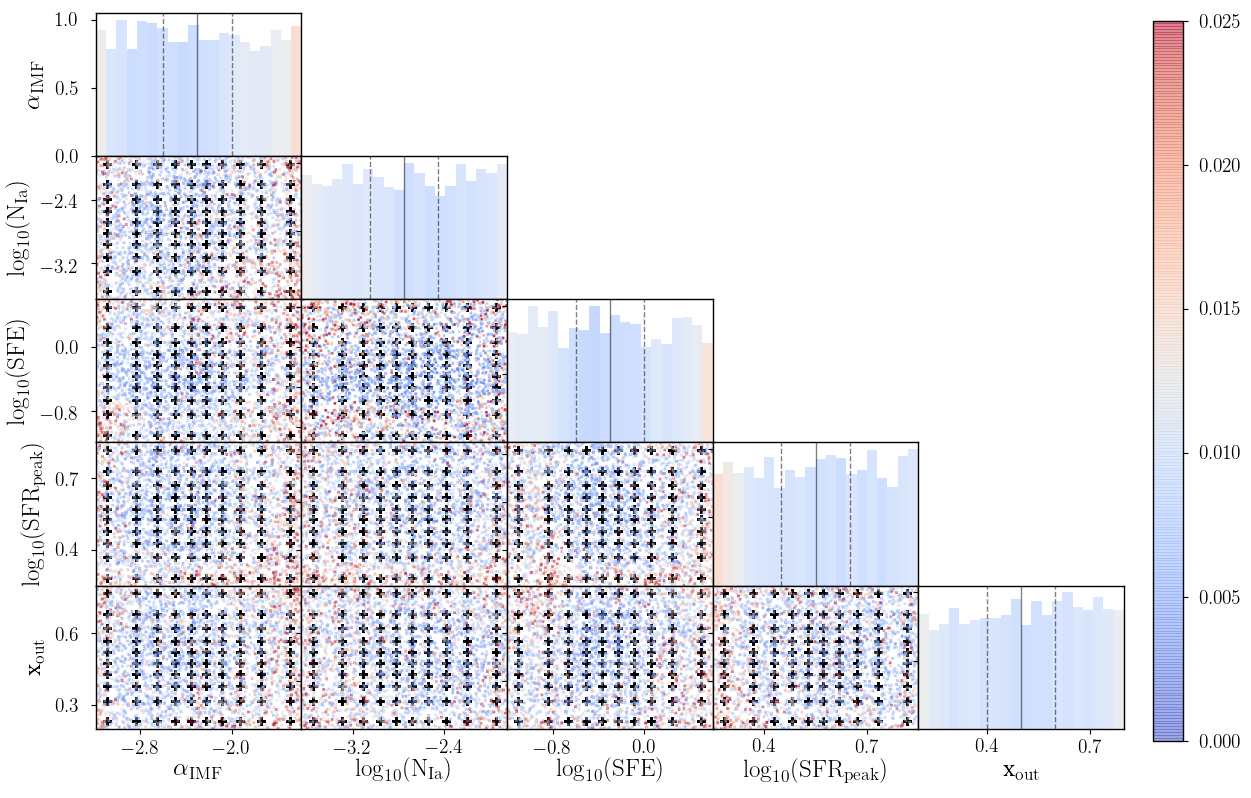}]\label{fig:NeuralErrors}
\end{figure}

To test the accuracy of the neural network, additional datasets were constructed with each parameter drawn from a uniform distribution of $3\sigma_\mathrm{prior}$ width centred on the prior mean. Figure \ref{fig:NeuralErrors} shows a corner plot of the five dimensional \textit{Chempy} parameter space, $\theta$, with coloured points representing the median L1 error between the neural network and \textit{Chempy} at that location in the parameter space slice. This is done for $n_\mathrm{el}=28$ and N13 yields, and histograms of the test dataset are shown on the diagonal. Average neural network errors across the $3\sigma_\mathrm{prior}$ range were $0.009^{+0.008}_{-0.004}$, and the figure shows that in the central regions, where the posterior peak concentrates, the error is far smaller. Crucially, these errors are insignificant compared to those of the minimum observational error of 0.02 dex. The network errors can be further reduced by training for a greater number of epochs, although this requires additional computation time.

\section{Posterior distributions and scores for all runs}\label{appenB}
Table \ref{tab:ism_parameters} includes the results from table \ref{tab:posteriors}, as well as the ISM parameters and the maximum posterior value for the \textit{Chempy} runs previously described. In addition, we include TNG yield runs with M$_\mathrm{CC-SN,\,max}=40$\,M$_\odot$ (here denoted $\mathrm{TNG_{40}}$), which are more comparable to our five tested yield sets. The TNG$_{40}$ runs yield similar results to the previous M$_\mathrm{CC-SN,\,max}=100$\,$\mathrm{M_\odot}$ (TNG$_{100}$) runs but with slightly poorer scores and a less bottom-heavy IMF. The Bayesian evidence for both TNG runs is highest when only the IMF is left free and N$_\mathrm{Ia}$ is fixed to the fiducial value (which is also recovered when both SSP parameters are left free). This results in a higher score because the parameter space is collapsed along this axis at a favourable value. 

For TNG$_{100}$, even the maximum achieved posterior is higher in the lower dimensional parameter space, though one would expect that leaving an additional parameter free would result in a better fit. Because of the high dimensionality, the MCMC never samples the exact peak, which will be hit much more easily by a Markov chain that has to sample in a lower dimensional space (under the condition that the missing parameter is set close to optimal value). 


The ISM parameters mainly reproduce the priors, albeit with a small correlation with the IMF high mass slope. If it tends to be top-heavy, as in case (a), the SFE is slightly higher, the SFR peaks earlier and the outflow fraction is lower, all of which slightly increase the ISM metallicity. The correlation points in the other direction when the IMF becomes slightly bottom-heavy, as in case (b).
	\begin{table*}
		\begin{tiny}
		\begin{minipage}{\textwidth}
			\begin{center}
				\caption{Inferred \textit{Chempy} and model error parameters (16, 50 and 84 percentiles of $\theta_\mathrm{posterior}$) obtained from an MCMC analysis for each yield set. The parameter priors and the maximum posterior obtained are also given. Results are given for: (a) the set of all available elements with $n_\mathrm{el}=28$; (b) the {\sc AREPO} most abundant element subset ($n_\mathrm{el}=8$); (c) the predicted parameters for the \citetalias{2017arXiv171004222G} simulations, using \citet[table 2]{Pillepich} TNG yields with M$_\mathrm{CC-SN,\,max}=100$\,M$_\odot$ (the default of IllustrisTNG) for five combinations of optimised SSP parameters; (d) following (c) but using M$_\mathrm{CC-SN,\,max}=40$\,M$_\odot$, matching this work.}
				\begin{tabular}{c|c|c|c||cc|cc|cc|cc|cc|cc}
					Yield set&\multicolumn{3}{c||}{Scores}&\multicolumn{2}{c|}{Error parameter}&\multicolumn{4}{c|}{SSP parameters}&\multicolumn{6}{c}{ISM parameters}\\
\hline

					\hline
					$\chi$&$\log_{10}(\mathcal{S}_\mathrm{B})$&$\log_{10}(\mathcal{S}_\mathrm{CV})$&Posterior&\multicolumn{2}{c|}{$\log_{10}\beta$}&\multicolumn{2}{c|}{$\alpha_\mathrm{IMF}$}&\multicolumn{2}{c|}{$\log_{10}(\mathrm{N_{Ia}})$}&\multicolumn{2}{c|}{$\log_{10}(\mathrm{SFE})$}&\multicolumn{2}{c|}{$\log_{10}(\mathrm{SFR_{peak}})$}&\multicolumn{2}{c}{$\mathrm{x_{out}}$}\\
					&$\mu$&$\mu$&$\log_{10}(\mathrm{P}_\mathrm{max})$&$\mu$&$\sigma$&$\mu$&$\sigma$&$\mu$&$\sigma$ &$\mu$&$\sigma$&$\mu$&$\sigma$&$\mu$&$\sigma$\\
					\hline
					&&&Priors:&1.0&$\pm0.5$&$-2.3$&$\pm0.3$&$-2.75$&$\pm0.3$&$-0.3$&$\pm0.3$&0.55&$\pm0.10$&0.5&$\pm0.1$\\
					\hline
					\multicolumn{16}{c}{\textit{(a) All 29 available elements}}\\
					\hline
					C04&$-1.21$&$-1.01$&$3.39$&0.77&$^{+0.30}_{-0.35}$&$-2.30$&$^{+0.11}_{-0.08}$&$-3.14$&$^{+0.16}_{-0.17}$&-0.24&$^{+0.28}_{-0.24}$&0.57&$^{+0.10}_{-0.09}$&0.48&$^{+0.09}_{-0.08}$\\
					N13 &$-5.69$&$-2.01$& $-7.35$ & 0.58 & $\pm0.29$ & $-2.32$ & $^{+0.16}_{-0.15}$ & $-2.90$ & $^{+0.18}_{-0.19}$&-0.31&$^{+0.30}_{-0.26}$&0.57&$^{+0.09}_{-0.11}$&0.51&$^{+0.11}_{-0.10}$\\
					W17&$-0.78$&$-1.01$&3.91&$0.72$&$^{+0.32}_{-0.36}$&$-1.83$&$^{+0.18}_{-0.22}$&$-3.29$&$^{+0.15}_{-0.18}$&$-0.27$&$^{+0.27}_{-0.17}$&$0.52$&$^{+0.11}_{-0.09}$&$0.44$&$^{+0.08}_{-0.09}$\\
					R17&$-6.11$&$-1.62$&$-9.18$&$0.50$&$^{+0.26}_{-0.36}$&$-2.29$&$^{+0.19}_{-0.23}$&$-2.91$&$^{+0.21}_{-0.23}$&$-0.20$&$^{+0.27}_{-0.25}$&$0.54$&$^{+0.11}_{-0.08}$&$0.46$&$^{+0.10}_{-0.11}$\\
P18&0.86&$-0.90$&7.36&$0.79$&$^{+0.27}_{-0.33}$&$-2.13$&$^{+0.21}_{-0.20}$&$-2.93$&$\pm0.16$&$-0.44$&$^{+0.30}_{-0.24}$&$0.59$&$\pm0.09$&$0.51$&$^{+0.10}_{-0.11}$\\
					\hline
					\multicolumn{16}{c}{\textit{(b) 9 most abundant elements}}\\
					\hline
					C04 &1.61&0.23&$10.53$&$1.01$&$^{+0.32}_{-0.35}$&$-2.45$&$^{+0.15}_{-0.11}$&$-2.89$&$^{+0.13}_{-0.12}$&$-0.45$&$^{+0.31}_{-0.26}$&$0.56$&$\pm0.09$&$0.52$&$\pm0.10$\\
					N13 &0.65&0.02&$7.99$&$0.83$&$^{+0.37}_{-0.36}$&$-2.52$&$^{+0.12}_{-0.11}$&$-2.79$&$^{+0.11}_{-0.12}$&$-0.44$&$^{+0.25}_{-0.22}$&$0.55$&$\pm0.09$&$0.53$&$^{+0.09}_{-0.12}$\\
					W17&0.73&$-0.18$&7.26&$0.90$&$^{+0.27}_{-0.30}$&$-2.19$&$^{+0.24}_{-0.17}$&$-3.09$&$^{+0.14}_{-0.16}$&$-0.33$&$^{+0.30}_{-0.25}$&$0.58$&$^{+0.09}_{-0.11}$&$0.51$&$\pm0.09$\\
					R17&$-0.49$&$-0.40$&3.44&$0.78$&$^{+0.31}_{-0.29}$
					&$-2.00$&$^{+0.23}_{-0.22}$&$-3.22$&$^{+0.21}_{-0.21}$&$-0.35$&$^{+0.37}_{-0.34}$&$0.57$&$\pm0.10$&$0.53$&$^{+0.11}_{-0.09}$\\
P18&$-0.01$&$-0.68$&5.25&0.72&$^{+0.35}_{-0.36}$&$-2.22$&$^{+0.20}_{-0.19}$&-2.84&$^{+0.16}_{-0.18}$&$-0.33$&$^{+0.27}_{-0.32}$&0.58&$^{+0.09}_{-0.10}$&0.52&$\pm0.10$\\
\hline
					\multicolumn{16}{c}{\textit{(c) like (b) but for TNG yields with} M$_\mathrm{CC-SN,max}=100$\,M$_\odot$}\\
                    \hline
                    $\mathrm{TNG_{fiducial}}$ &$-0.89$&$-0.98$&1.54&$0.64$&$\pm0.33$&$-2.3$&fixed&$-2.89$&fixed&$-0.12$&$^{+0.26}_{-0.24}$&$0.52$&$^{+0.10}_{-0.09}$&$0.52$&$^{+0.08}_{-0.10}$\\
                    $\mathrm{TNG_{SN\,Ia\,free}}$ &$-0.31$&$-0.15$&3.41&$0.80$&$^{+0.29}_{-0.36}$&$-2.3$&fixed&$-2.63$&$\pm0.13$&$-0.32$&$^{+0.31}_{-0.34}$&$0.58$&$^{+0.09}_{-0.11}$&$0.55$&$^{+0.09}_{-0.11}$\\
                    $\mathrm{TNG_{IMF\,free}}$ &2.12&0.34&12.26&$1.17$&$^{+0.41}_{-0.43}$&$-2.72$&$\pm0.05$&$-2.89$&fixed&$-0.34$&$^{+0.21}_{-0.14}$&$0.49$&$\pm0.06$&$0.44$&$^{+0.10}_{-0.08}$\\
                    $\mathrm{TNG_{IMF\,\&\,SN\,Ia\,free}}$ &1.82&0.29&12.01&$1.13$&$\pm0.46$&$-2.68$&$^{+0.08}_{-0.09}$&$-2.87$&$^{+0.10}_{-0.09}$&$-0.28$&$^{+0.24}_{-0.18}$&$0.50$&$^{+0.10}_{-0.09}$&$0.47$&$\pm0.10$\\
					\hline
                    \multicolumn{16}{c}{\textit{(d) like (c) but with} M$_\mathrm{CC-SN,max}=40$\,M$_\odot$}\\
                    \hline
					$\mathrm{TNG_{fiducial}}$  &$-0.40$&$-0.69$&2.81&$0.75$&$^{+0.26}_{-0.37}$ &$-2.3$&fixed&$-2.89$&fixed&$-0.31$&$^{+0.29}_{-0.19}$&$0.53$&$^{+0.09}_{-0.08}$&$0.51$&$\pm0.09$\\
					$\mathrm{TNG_{SN\,Ia\,free}}$ &$-0.45$&$-0.20$&4.00&$0.76$&$^{+0.32}_{-0.38}$ &$-2.3$&fixed&$-2.69$&$^{+0.12}_{-0.15}$&$-0.55$&$^{+0.34}_{-0.23}$&$0.59$&$^{+0.08}_{-0.10}$&$0.53$&$\pm0.10$\\
					$\mathrm{TNG_{IMF\,free}}$ &1.69&0.20&11.18&$1.02$&$^{+0.41}_{-0.37}$ &$-2.64$&$\pm0.06$&$-2.89$&fixed&$-0.41$&$^{+0.20}_{-0.11}$&$0.50$&$^{+0.08}_{-0.07}$&$0.46$&$\pm0.09$\\
					$\mathrm{TNG_{IMF\,\&\,SN\,Ia\,free}}$ &1.19&0.18&11.41&$1.07$&$^{+0.36}_{-0.42}$ &$-2.64$&$^{+0.10}_{-0.08}$&$-2.90$&$^{+0.13}_{-0.11}$&$-0.41$&$^{+0.18}_{-0.15}$&$0.48$&$\pm0.09$&$0.47$&$\pm0.09$
                    \label{tab:ism_parameters}
				\end{tabular}
			\end{center}
		\end{minipage}
	\end{tiny}
	\end{table*}




\end{document}